Chapter

# Perspective Chapter: Insights from Kalman Filtering with Correlated Noises Recursive Least-Square Algorithm for State and Parameter Estimation

*Abd El Mageed Hag Elamin Khalid*

## Abstract

This article explores the estimation of parameters and states for linear stochastic systems with deterministic control inputs. It introduces a novel Kalman filtering approach called *Kalman Filtering with Correlated Noises Recursive Generalized Extended Least Squares (KF-CN-RGELS) algorithm*, which leverages the cross-correlation between process noise and measurement noise in Kalman filtering cycles to jointly estimate both parameters and system states. The study also investigates the theoretical implications of the correlation coefficient on estimation accuracy through performance analysis involving various correlation coefficients between process and measurement noises. The research establishes a clear relationship: the accuracy of identified parameters and states is directly proportional to positive correlation coefficients. To validate the efficacy of this algorithm, a comprehensive comparison is conducted among different algorithms, including the standard Kalman filter algorithm and the augmented-state Kalman filter with correlated noises algorithm. Theoretical findings are not only presented but also exemplified through a numerical case study to provide valuable insights into practical implications. This work contributes to enhancing estimation accuracy in linear stochastic systems with deterministic control inputs, offering valuable insights for control system design and state-space modeling.

## 1. Introduction

Finding a state-space model is crucial for designing control systems. A big part of this process involves changing the unknown parameters in a controller's transfer function or state-space representation to get the desired stability, performance, and robustness [1]. In the field of control, system identification has been done using many different estimation methods, such as least-squares methods [2], iterative identification methods [3], Bayesian methods [4, 5], separated least-squares methods [6, 7], and maximum likelihood methods [8]. However, identifying the parameters of state-space models is more difficult. For instance, a state-space model has unknown states as well as unknown parameter matrices or vectors [9, 10]. In this area, some studies focus on identifying nonlinear state-space models [11]. Meanwhile, other studies look at how to identify linear state-space models. For example, Safarinejadian et al. [12] introduced a new way to identify a state-space model for a single input single-output fractional-order system based on a new fractional-order Kalman filter with correlated noises. To find the parameters of linear state-space models, Yu et al. [13] used the least-squares estimation framework and made the Hankel matrix factorization less affected by Markov-parameter estimation error by using a single optimization framework instead of the two-step method used by Yu et al. [14] to identify the structured system matrices. A study [8] introduced two hidden variables and used the expectation-maximization algorithm to estimate time delays and parameters in a state-space model with unknown time delay. To identify slowly changing linear time-varying systems, Razmjooei et al. [15] used a two-step method: first, they estimated parameters using Legendre basis functions, then used the Kalman filter with those estimated parameters to compute the system's states. Industrial processes often suffer from noise in both measurements and the processes themselves, making accurate state and parameter estimation challenging. To address this, Li et al. [16] applied data filtering to mitigate the impact of colored noise in the measurement equation for bilinear state-space models. Recognizing the influence of both process and measurement noise, Cui et al. [17] leverages data filtering to mitigate the effects of colored noise, specifically within the state equations. Aiming to reduce the impact of colored noise and achieve better estimation accuracy, Wang et al. [18] proposed a novel algorithm. This approach combines filtering techniques with a recursive generalized least-squares method, utilizing filtered measurement data for continuous updates [19]. Ma et al. [20] focused on identifying state-space models in a specific form (observer canonical) while dealing with white noise affecting the output measurements. Building upon existing methods, Cui et al. [21] presented a novel algorithm for jointly estimating parameters and states in a specific class of state-space models (observer canonical) that are exposed to colored noise. This approach leverages both Kalman filtering and gradient search techniques for improved accuracy. In Cui et al. [22], researchers tackle a specific noise situation (white noise in state equations, moving average noise in measurements) by developing two algorithms for jointly estimating states and parameters in state-space systems. These algorithms rely on the "auxiliary model" concept. Motivated by the limitations of existing methods, this study investigates state-space models experiencing specific noise characteristics (white noise, moving average noise) and correlated noise behaviors, aiming to develop robust estimation algorithms for such systems. In the field of parameter and state estimation, researchers are striving to achieve both more accurate results and faster calculations. Multi-innovation identification theory tackles this challenge by utilizing more system and data information to improve parameter estimation accuracy, as shown in previous studies [23–28]. Building upon past studies of correlated noise, this work takes a deeper dive into the specific effects of varying correlation coefficients on estimation accuracy. This knowledge will inform the development of improved estimation algorithms for state-space systems under realistic noise conditions.

This paper makes four main contributions:

1. The paper challenges the conventional assumption of the Kalman filter, which relies on known system parameters. It uses past estimates of the system's parameters to predict its current state and then uses those predictions to improve the estimates of those parameters.

2. In the context of correlated process noise and measurement noise, the paper introduces a new formulation of the Kalman filter algorithm that preserves the fundamental assumption of Kalman filtering, which operates with uncorrelated noise.

3. Emphasizing the significant role of the correlation coefficient between process noise and measurement noise, the paper demonstrates that higher correlation coefficients lead to more accurate estimates.

4. The negative correlation coefficient reduces the estimation and identification accuracy, considering accuracy from two different perspectives: observation and model. It leads to an increase in measurement covariance and cross-covariance between state and process noise. This increase in covariance and cross-covariance affects the filtering results, leading to less accurate state estimates and parameter estimation.

This article is structured as follows. Section 2 describes the system model of a linear stochastic state-space system. Section 3 contains the formulation of the Kalman filter for handling cross-correlated noise. Section 4 presents the algorithm used for the comparative evaluation. Section 5 demonstrates the formulation of the identification model for linear stochastic state-space models. Section 6 contains the derivation of the proposed algorithm (KF-CN-RGELS). Section 7 explains the impact of negative correlation coefficients. Section 8 provides an example to verify the effectiveness of the proposed algorithm. Concluding observations can be found in Section 9.

## 2. The system model for linear stochastic system

Let us introduce some notation. The expressions "$F=:X$" or "$X:=F$" indicate that "$F$" is defined as "$X$." The symbol "$q$" represents a unit back-shift operator, where $q^{-1}x(t)$ denotes $x(t-1)$. The superscript "$T$" denotes the transpose of vectors/matrices [29].

Consider the following linear-stochastic system:

$$x(t+1) = Fx(t) + Gu(t) + w(t), \tag{1}$$

$$\mathrm{y}(t) = Hx(t) + du(t) + v^*(t), \tag{2}$$

$$v^*(t) := \left(1 + J_1 q^{-1} + J_2 q^{-2} + \cdots + J_{n_J} q^{-n_J}\right) v(t), \tag{3}$$

Here, $u(t) \in \mathbb{R}$ and $y(t) \in \mathbb{R}$ are the input and output of the system, respectively. The state vector of the system is denoted as $x(t) := [x_1(t), \ldots, x_n(t)]^T \in \mathbb{R}^n$. The white noise process $v(t) \in \mathbb{R}$ has zero mean and variance $\sigma_v^2$, and $w(t) := [w_1(t), \ldots, w_n(t)]^T \in \mathbb{R}^n$ represents the white noise vector with zero mean. $J(q)$ in the unit back-shift operator $q^{-1}$ is defined as $J(q) = 1 + J_1 q^{-1} + J_2 q^{-2} + \cdots + J_{n_J} q^{-n_J} \in \mathbb{R}$. The matrices $F$, $G$, and $H$ representing system parameters are defined as follows:

$$F := \begin{bmatrix} -f_1 & 1 & 0 & \cdots & 0 \\ -f_2 & 0 & 1 & \cdots & 0 \\ \vdots & \vdots & \vdots & \ddots & \vdots \\ -f_{n-1} & 0 & 0 & \cdots & 1 \\ -f_n & 0 & 0 & \cdots & 0 \end{bmatrix} \in \mathbb{R}^{n\times n}, G := \begin{bmatrix} g_1 \\ g_2 \\ \vdots \\ g_{n-1} \\ g_n \end{bmatrix} \in \mathbb{R}^n, H := [1, \ 0, \ \cdots \ 0] \in \mathbb{R}^{1\times n} \quad (4)$$

## 3. Formulation of Kalman filter with noises cross-correlation

In this section, we delve into the formulation of a Kalman filter designed to address the presence of noise cross-correlation in linear stochastic state-space models. We make several key assumptions and present the mathematical derivations essential for a clear understanding of this formulation.

**Assumptions:** Consider a linear stochastic state-space model described by Eqs. (1) and (2). In this framework:

- The sequences $w(t)$ and $v(t)$ are zero-mean Gaussian white noise processes.
- The variance of $w(t)$ at time $t$ is represented by $Q(t)$, which is a positive definite matrix.
- The variance of $v(t)$ at time $t$ is denoted as $R(t)$, also a positive definite matrix.
- These noise sequences, $w(t)$ and $v(t)$, exhibit statistical correlation. Specifically, the correlation is defined by equation [30, 31]:

$$E\left[w(k)v(l)^T\right] = S(k)\delta_{kl}, , k, l = 0, 1, \ldots \quad (5)$$

where $\delta_{kl}$ represents the Kronecker delta function, and each $S(k)$ is a non-negative definite matrix.

**Formulation:** To effectively deal with correlated noise sequences, a gain matrix $T$ is introduced and incorporated into the system equations. This process involves adding the gain matrix $T$ to the measurement Eq. (2). The modified system Eq. (1) now reads:

$$x(t+1) \quad = Fx(t) + Gu(t) + w(t) + T[y(t) - Hx(t) - du(t) - J(q)v(t)] \quad (6)$$

To simplify the representation, a noise term $v^*(t)$ is defined as follows:

$$v^*(t) := \left(1 + J_1 q^{-1} + J_2 q^{-2} + \cdots + J_{n_J} q^{-n_J}\right) v(t) \quad (7)$$

As a result, the modified system equation becomes:

$$x(t+1) \quad = (F - TH)x(t) + (G - Td)u(t) + Ty(t) + [w(t) - Tv^*(t)]. \quad (8)$$

In Eq. (8), the term $Ty(t)$ represents a determined control input. To further simplify, we introduce the following notations:

$$\overline{F}:=(F - TH), \tag{9}$$

$$\overline{G}:=(G - Td), \tag{10}$$

$$\overline{w}(t):=w(t) - Tv^*(t) \tag{11}$$

With these notations, the modified system dynamics equation can be succinctly rewritten as:

$$x(t+1) = \overline{F}x(t) + \overline{G}u(t) + Ty(t) + \overline{w}(t). \tag{12}$$

### 3.1 Derivation of the gain matrix *T*

From Eq. (11), the mean and covariance of $\overline{w}(t)$ can be described as follows:

$$E[\overline{w}(t)] = E[w(t)] - TE[v^*(t)] = 0, \tag{13}$$

$$E\left[\overline{w}(t)\overline{w}^T(t)\right] = \overline{Q}. \tag{14}$$

Notably, we have assumed that the noises are correlated at the same time. By employing Eq. (7), we can demonstrate the relationship:

$$E\left[w(t)v^T(t)\right] = E\left[w(t)v^{*T}(t)\right] = S. \tag{15}$$

Substituting Eq. (11) into (14), we obtain the following expression for $\overline{Q}$:

$$\overline{Q} = Q + (TR - S)T^T - TS^T. \tag{16}$$

Taking the transpose of Eq. (11), we define:

$$\overline{w}^T(t) = w^T(t) - v^{*T}(t)T^T \tag{17}$$

When we multiply Eq. (11) by $\overline{w}^T(t)$ and take the expectation, we get:

$$E\left[\overline{w}(t)\overline{w}^T(t)\right] = E\left[\overline{w}(t)w^T(t)\right] - E\left[\overline{w}(t)v^{*T}(t)\right]T^T \tag{18}$$

In line with the standard Kalman filter assumptions, we assume that the new process noise, $\overline{w}(t)$, is independent of the measurement noise $v^*(t)$, meaning $E\left[\overline{w}(t)v^{*T}(t)\right] = 0$. With this assumption, we arrive at the relationship:

$$E\left[\overline{w}(t)\overline{w}^T(t)\right] = E\left[\overline{w}(t)w^T(t)\right] = \overline{Q} \tag{19}$$

By right-multiplying Eq. (11) by $w^T(t)$ and taking the expected value, we can establish:

$$E\left[\overline{w}(t)w^T(t)\right] = E\left[w(t)w^T(t) - TE\left[w^T(t)v^*(t)\right]\right. \tag{20}$$

This relationship leads to the equation:

$$\overline{Q} = Q - TS^T \tag{21}$$

By subtracting Eq. (21) from (16), we can derive the following relationship:

$$TR - S = 0, \text{ since } T \neq 0 \tag{22}$$

Hence, the gain matrix $T$ needed to handle the uncorrelated process noise $\overline{w}(t)$ and the measurement noise $v^*(t)$ can be calculated as:

$$T = SR^{-1} \tag{23}$$

**Remark 1:** It is crucial to consider the new variance $\overline{Q} = Q - TS^T$ of the noise sequence $\overline{w}(t)$. This is a critical adjustment to the model, as it departs from the variance $Q$ of the noise sequence $w(t)$.

### 3.2 Kalman filter prediction and correction cycles

At this point, we focus on the system dynamics equation and the measurement equation, which are given in Eqs. (24)–(25):

$$x(t+1) = \overline{F}x(t) + \overline{G}\,u(t) + Ty(t) + \overline{w}(t) \tag{24}$$

$$y(t) = Hx(t) + du(t) + v^*(t) \tag{25}$$

The prediction and correction cycles of the modified Kalman filter for this proposed system are as follows:

**Prediction cycle**

For $x_p(t)$ representing the prediction state, the prediction cycle equations are:

$$x_p(t+1) = \overline{F}x_p(t) + \overline{G}\,u(t) + Ty(t) \tag{26}$$

$$P_p = \overline{F}P_p\overline{F}^T + \overline{Q} \tag{27}$$

$$K = P_pH^T\left(HP_pH^T + R\right)^{-1} \tag{28}$$

Here, $K$ and $P_p$ represent the Kalman gain and the estimation error covariance, respectively.

**Correction cycle**

For $x_c(t)$ be the corrected state or the estimated state, the correction cycle equations will be as follows.

$$x_c(t+1) = x_p(t+1) + Kv^*(t) \tag{29}$$

(where) $v^*(t) = y(t) - c * x(t) - d * u(t)$

The estimated error covariance is calculated as:

$$P_c = (I_n - KH)P_p, \tag{30}$$

Eqs. (26)–(30) describe the prediction and correction cycles of the Kalman filtering process when dealing with correlated noise sequences.

## 4. Introduction and formulation of Kalman filtering algorithms for comparative assessment

In this section, we introduce and formulate two Kalman filtering algorithms for the purpose of conducting a comprehensive comparative analysis with the proposed algorithm. The algorithms to be compared are the standard Kalman filter (SKF) and the State Augmented Kalman Filter with Correlated Noises (AUG-KF). Each of these algorithms is presented with its assumptions and key equations for prediction and correction cycles [32–34].

### 4.1 Standard Kalman filter (SKF)

Consider the system model described by Eqs. (1)–(2)

**Assumptions:** The process noise ($w(t)$) and measurement noise ($v(t)$) are independent.

**Formulation:** Prediction and Correction Cycles

**Prediction cycle**

Predicted state: $x_p(t+1) = Fx_p(t) + G\,u(t)$

Prediction error covariance: $P_p = FP_pF^T + Q$

Kalman gain: $K = \left(P_pH^T\right)\left(HP_pH^T + R\right)^{-1}$

**Correction cycle**

Corrected state: $x_c(t+1) = x_p(t+1) + Kv^*(t)$, Where $v^*(t) = y(t) - c * x(t) - d * u(t)$.

Correction error covariance: $P_c = P_p - K\left(P_pH^T\right)^T$

### 4.2 Augmented-State Kalman filter (AUG-KF)

State augmentation is a technique to extend the state vector of a system by adding some auxiliary variables that are related to the original state or the noise. In [35], the authors augment the state vector by appending the process noise vector, so that the cross-correlated noises can be treated as independent noises in the augmented system. This technique can simplify the filter design and improve the estimation accuracy, but it also increases the dimension of the state vector and the computational cost of the filter. With reference to method used in [35] we can derive the AUG-KF algorithm for the system (1)–(2) as

**Assumptions:** The process noise ($w(t)$) and measurement noise ($v(t)$) are correlated with $E\left[w(t)v^*(t)^T\right] = S$ represents the cross-covariance between process noise $w(t)$ and measurement noise $v^*(t)$.

**Formulation:** Initialization, Prediction and Update Cycles:

**Initialization:**

$n_t = y(t) - H\,x(t) - d\,u(t) - J_1\,v(t-1) - J_2\,v(t-2)$, where $n_t$ is the actual observation.

**Augmented Kalman Gains:**

$k_a = [k_x, k_w]$, where $k_x = P\,H^T\left(HP_pH^T + R\right)^{-1}$, $k_w = S\,\left(HP_pH^T + R\right)^{-1}$.

Update process noise and state estimate:

$w(t) = k_w\;n_t$, $x_p(t+1) = Fx_p(t) + G\,u(t) + w(t)$.

$$P_p(\mathrm{t}+1) = F\,P_p(\mathrm{t})F^T + P_w + F\,P_{xw} + P_{wx}\,F^T$$

**Update covariance matrix of augmented state and process noise:**
$P_{Aug} = \begin{bmatrix} P_c(\mathrm{t}) & P_{xw} \\ P_{wx} & P_w \end{bmatrix}$, where $P_c(\mathrm{t}) = P_p(\mathrm{t}+1) - k_x\,\left(HP_p(t+1)H^T + R\right)^{-1} {k_x}^T$,
$P_w = Q - k_w\,\left(HP_c(t)H^T + R\right)^{-1} {k_w}^T$, $P_{wx} = -k_w\,\left(HP_c(t)H^T + R\right)^{-1} {k_x}^T$
And $P_{xw} = -k_x\left(HP_c(t)H^T + R\right)^{-1} {k_w}^T$.
**Update for the next time step:**
Corrected state:

$$x_c(\,t+1) = x_p(t+1) + k_x\;\; n_t,$$

Correction error covariance:

$$P_c(t+1) = P_c(t) - k_x\,\left(HP_c(t)H^T + R\right)^{-1} {k_x}^T.$$

These algorithms will form the foundation for a comparative analysis in the following sections. We will assess their performance and effectiveness in managing noise correlations, comparing them to the proposed algorithm. This evaluation aims to identify the most effective approach in various scenarios.

## 5. The identification model for linear stochastic system

Let us define some essential notation:

- "$I_n$" denotes an identity matrix of appropriate size, typically $n \times n$.
- "$\mathbf{1}_n$" represents an $n$-dimensional column vector with all elements equal to unity.
- "$\hat{\theta}(t)$ "represents the estimate of $\theta$ at time $t$, and "$\hat{x}(t)$" denotes the estimate of $x(t)$.

Now, based on Section 3, we modify the system equations:

$$x(t+1) = \overline{F}x(t) + \overline{G}u(t) + Ty(t) + \overline{w}(t) \tag{31}$$

$$y(t) = Hx(t) + du(t) + v^*(t) \tag{32}$$

With

$$v^*(t) := \left(1 + J_1q^{-1} + J_2q^{-2} + \cdots + J_{n_J}q^{-n_J}\right)v(t)$$

$\overline{F} := (F - TH)$, $\overline{w}(t) := w(t) - Tv^*(t)$.

The system's input and output are represented by $u(t) \in R$ and $y(t) \in R$, respectively, with $x(t) := [x_1(t),\ \dots, x_n(t)]^T \in \mathbb{R}^n$ represents the system state vector. $v(t) \in R$ is a white noise process with zero mean and variance $\sigma_v^2$, and $w(t) := [w_1(t),\ \dots, w_n(t)]^T \in \mathbb{R}^n$ denotes the white noise vector with zero mean. The polynomial $J(q)$ in the unit backshift operator $q^{-1}$ is expressed as $J(q) = 1 + J_1 q^{-1} + J_2 q^{-2} + \cdots + J_{n_J} q^{-n_J} \in \mathbb{R}$. We also have system parameters: $F \in \mathbb{R}^{n\times n}, G \in \mathbb{R}^n, H \in \mathbb{R}^{1\times n}$ and $d \in (\mathbb{R})$

The matrices $F$, $G$ and $H$ are defined as follows:

$$\overline{F} := \begin{bmatrix} -\overline{f_1} & 1 & 0 & \cdots & 0 \\ -\overline{f_2} & 0 & 1 & \cdots & 0 \\ \vdots & \vdots & \vdots & \ddots & \vdots \\ -\overline{f}_{n-1} & 0 & 0 & \cdots & 1 \\ -\overline{f}_n & 0 & 0 & \cdots & 0 \end{bmatrix} \in \mathbb{R}^{n\times n},$$
$$\overline{G} := \begin{bmatrix} \overline{g}_1 \\ \overline{g}_2 \\ \vdots \\ \overline{g}_{n-1} \\ \overline{g}_n \end{bmatrix} \in \mathbb{R}^n, H := [1, \quad 0, \quad \cdots \quad 0] \in \mathbb{R}^{1\times n} \tag{33}$$

Assuming that $y(t), u(t), w(t)$, and $v(t)$ are strictly proper, meaning their values are zero for 0 for $t \le 0$, and that the orders $n$ and $n_J$ are known, we can derive the following state equations from (1)–(3):

$$\begin{aligned} & x_1(t) = -\overline{f_1} x_1(t-1) + x_2(t-1) + \overline{g}_1 u(t-1) \\ & \quad + \overline{w_1}(t-1) + T_1 y(t-1) \\ & x_2(t-1) = -\overline{f}_2 x_1(t-2) + x_3(t-2) + \overline{g}_2 u(t-2) \\ & \quad + \overline{w_2}(t-2) + T_2 y(t-2) \\ & x_3(t-2) = -\overline{f}_3 x_1(t-3) + x_4(t-3) + \overline{g}_3 u(t-3) \\ & \quad + \overline{w_3}(t-3) + T_3 y(t-3) \\ & \vdots \\ & x_{n-1}(t-n+2) = -\overline{f}_{n-1} x_1(t-n+1) + x_n(t-n+1) \\ & \quad + \overline{g}_{n-1} u(t-n+1) + \overline{w}_{n-1}(t-n+1) + T_{n-1} y(t-n+1) \\ & x_n(t+1-n) = -\overline{f}_n x_1(t-n) + \overline{g}_n u(t-n) + \overline{w}_n(t-n) + T_n y(t-n). \end{aligned} \tag{34}$$

From these $n$ equations, we can express $x_1(t)$ as:

$$x_1(t) = -\sum_{i=1}^{n} \overline{f}_i x_1(t-i) + \sum_{i=1}^{n} \overline{g}_i u(t-i) + \sum_{i=1}^{n} \overline{w}_i(t-i) + \sum_{i=1}^{n} T_i y(t-i) \tag{35}$$

Substituting $H = [1, 0, \ldots, 0]$ in (2), we obtain:

$$y(t) = x_1(t) + du(t) + \left[1 + J_1 q^{-1} + \cdots + J_{n_J} q^{-n_J}\right] v(t) \tag{36}$$

Now, let us define the parameter vector $\theta$ and the information vector $\varphi(t)$:

$$\overline{\theta}:=\left[\overline{F}^T, \overline{G}^T, d, J^T\right]^T \in \mathbb{R}^{n_s}, n_s = 2n + 1 + n_J, \quad (37)$$

$$\overline{F}:=\left[\overline{f}_1, \overline{f}_2, \dots, \overline{f}_n\right]^T \in \mathbb{R}^n, \quad (38)$$

$$J:=\left[J_1, J_2, \dots, J_{n_J}\right]^T \in \mathbb{R}^{n_J}, \quad (39)$$

$$\varphi(t):=\left[\varnothing_{\overline{f}}(t)^T, \varnothing_{\overline{g}}(t)^T, u(t), \varnothing_v(t)^T\right]^T \in \mathbb{R}^{n_s} \quad (40)$$

Where:

$$\varnothing_{\overline{f}}(t):=[-x_1(t-1), \dots, -x_1(t-n)]^T \in \mathbb{R}^n$$

$$\varnothing_{\overline{g}}(t):=[u(t-1), u(t-2), \dots, u(t-n)]^T \in \mathbb{R}^n$$

$$\varnothing_v(t):=\left[v(t-1), v(t-2), \dots, v\left(t-n_J\right)\right]^T \in \mathbb{R}^{n_J}$$

Now, we introduce the variables $\gamma(t)$, and $\beta(t)$ defined as:

$$\gamma(t):=\sum_{i=1}^{n} w_i(t-i) = w_1(t-1) + w_2(t-2) + \dots + w_n(t-n) \quad (41)$$

$$\beta(t):=\sum_{i=1}^{n} T_i y(t-i) = T_1 y(t-1) + T_2 y(t-2) + \dots + T_n y(t-n) \quad (42)$$

Using (35) and (41), Eq. (36) can be rewritten as:

$$\begin{aligned} y(t) &= \varnothing_{\overline{f}}(t)^T \overline{F} + \varnothing_{\overline{g}}(t)^T \overline{G} + \gamma(t) + \beta(t) + u(t) + \varnothing_v(t)^T J + v(t) \\ &= \varphi(t)^T \theta + \gamma(t) + \beta(t) + v(t) \end{aligned} \quad (43)$$

Eq. (43) represents the identification model for a linear stochastic state-space system as defined in (1)–(2). The main objective of this paper is to present an algorithm that jointly estimates the system states and unknown parameters using recursive generalized extended least squares. Additionally, we aim to investigate the impact of correlations between process and measurement noises on estimation accuracy.

**Remark 2:** To simplify the identification process, the observable general state-space system described in (1)–(2) is transformed into the observer canonical form. This transformation serves to reduce the number of parameters that need to be identified, making the estimation process more efficient and accurate.

## 6. The KF-CN-RGELS algorithm

This section introduces the KF-CN-RGELS algorithm for the joint estimation of system parameters and states in a canonical observer state-space system (Eqs. 1 and 2).

The algorithm comprises two key components: the parameter estimation algorithm and the state estimation algorithm. These algorithms are developed to address the challenge of estimating parameters and states in the proposed system, and when combined, they provide a comprehensive solution.

### 6.1 The parameter estimation algorithm

The parameter estimation algorithm is centered around minimizing the quadratic criterion function defined as:

$$C(\theta) := \sum_{j=1}^{L} \|y(j) - \varphi(j)^T\theta - \gamma(j) - \beta(j)\|^2, \tag{44}$$

This minimization process helps estimate the system parameters based on the identification model (Eq. 43) using the least-squares principle [22]. The parameter estimation is performed recursively, and it can be expressed as follows:

$$\hat{\theta}(t) = \hat{\theta}(t-1) + L(t)\left[y(t) - \gamma(t) - \beta(t) - \varphi(t)^T\hat{\theta}(t-1)\right] \tag{45}$$

$$L(t) = \frac{P(t-1)\varphi(t)}{1 + \varphi(t)^T P(t-1)\varphi(t)} \tag{46}$$

$$P(t) = P(t-1) - L(t)[P(t-1)\varphi(t)]^T, P(0) = p_0 I_n \tag{47}$$

However, in implementing these algorithms, certain challenges arise due to unknown process and measurement noise sequences, as well as unmeasurable states. To address these challenges, the concept of auxiliary models is introduced to replace unknown parameters and states with their estimates. In this modified parameter estimation algorithm, $\hat{\varphi}(t)$ and $\hat{\gamma}(t)$ are used instead of $\varphi(t)$ and $\gamma(t)$, resulting in the following algorithm [22]:

$$\hat{\theta}(t) = \hat{\theta}(t-1) + L(t)\left[y(t) - \hat{\gamma}(t) - \beta(t) - \hat{\varphi}(t)^T\hat{\theta}(t-1)\right] \tag{48}$$

$$L(t) = \frac{P(t-1)\hat{\varphi}(t)}{1 + \hat{\varphi}(t)^T P(t-1)\hat{\varphi}(t)} \tag{49}$$

$$P(t) = P(t-1) - L(t)[P(t-1)\hat{\varphi}(t)]^T, P(0) = p_0 I_n \tag{50}$$

The vectors $\hat{\phi}_v(t)$, $\hat{\gamma}(t)$ and $\beta(t)$ are formed using the output, the process noise, and measurement noise sequences, which are calculated using the equations provided.

To form $\hat{\phi}_v(t)$ in $\hat{\varphi}(t)$ and $\hat{\gamma}(t), \hat{w}(t-i)$ and, $\hat{v}(t-i)$ can be calculated using (1) and (2) as

$$\hat{\overline{w}}(t) = \hat{x}(t+1) - \hat{\overline{\overline{F}}}(t)\hat{x}(t) - \hat{\overline{G}}(t)u(t) - T\left(cx(t) + \hat{d}\, u(t)\right) \tag{51}$$

$$\widehat{v^*}(t) = y(t) - H\hat{x}(t) - \hat{d}(t)u(t), \tag{52}$$

$$\hat{v}(t) = \widehat{v^*}(t) - \hat{J}_1\hat{v}(t-1) - \hat{J}_2\hat{v}(t-2) \tag{53}$$

### 6.2 The state estimation algorithm

This section deals with the problem of estimating non-measurable states of the information vector $\varphi(t)$. To overcome this challenge, a modified Kalman filter prediction and update cycle, described in Section 4, is used to estimate the system state. The state estimation depends on the degree of correlation between process noise and measurement noise in the proposed system. The state estimation algorithm is as follows:

**Prediction cycle:**

$$xp(t+1) = \hat{\overline{F}}xp(t) + \hat{\overline{G}}u(t) + Ty(t); \tag{54}$$

$$P = \hat{\overline{F}}P\hat{\overline{F}}^T + Q - \hat{T}\hat{S}^T \tag{55}$$

**Calculation of Kalman gain:**

$$K = (Pc)/c\,Pc^T + \hat{R} \tag{56}$$

**Correction cycle:**

$$x(t+1) = xp(t+1) + +K\widehat{v^*}(t) \tag{57}$$

$$P = (eye(n) - Kc)P \tag{58}$$

In these equations, $\overline{F} = (F - TH), \overline{Q} = Q - TS^T$ and $S = \rho_{w,v}\sqrt{R}\sqrt{Q}$, as stated in Section 4. Therefore,

$$\hat{x}(t) = x_c(t) \tag{59}$$

where $x_p(t)$ and $x_c(t)$ are predicted and corrected states at time $t$, (respectively)

**Remark 3:** A Kalman filter estimates the state of a system, assuming that its parameters are known. To overcome this issue, the concept of an auxiliary model that replaces all system parameters with estimates in the prediction and correction cycles of the modified Kalman filter recursion equations, as illustrated in (55)–(60), is applied.

### 6.3 The joint parameter and state estimation algorithm

The joint parameter and state estimation algorithm combines the parameter estimation and state estimation algorithms to recursively estimate both the system's parameter vector and state vector. The algorithm is described by the following equations:

$$\hat{\theta}(t) = \hat{\theta}(t-1) + L(t)\left[y(t) - \hat{\gamma}(t) - \beta(t) - \hat{\varphi}(t)^T\hat{\theta}(t-1)\right] \tag{60}$$

$$L(t) = \frac{P(t-1)\hat{\varphi}(t)}{1 + \hat{\varphi}(t)^T P(t-1)\hat{\varphi}(t)} \tag{61}$$

$$P(t) = P(t-1) - L(t)[P(t-1)\hat{\varphi}(t)]^T, P(0) = p_0 I_n \tag{62}$$

$$\hat{\varphi}(t) = \left[\hat{\varnothing}_{\bar{f}}(t)^T, \varnothing_g(t)^T, u(t), \hat{\varnothing}_v(t)^T\right]^T \in \mathbb{R}^{n_s}, \tag{63}$$

$$\hat{\varnothing}_f(t) := [-\hat{x}_1(t-1), -\hat{x}_1(t-2), \ldots, -\hat{x}_1(t-n)]^T \in \mathbb{R}^n \tag{64}$$

$$\varnothing_b(t) := [u(t-1), u(t-2), \ldots, u(t-n)]^T \in \mathbb{R}^n \tag{65}$$

$$\hat{\varnothing}_v(t) := \left[\hat{v}(t-1), \hat{v}(t-2), \ldots, \hat{v}\left(t-n_J\right)\right]^T \in \mathbb{R}^{n_J} \tag{66}$$

$$\hat{\gamma}(t) = \hat{\overline{w}}(t-1) + \hat{\overline{w}}(t-2) + \cdots + \hat{\overline{w}}(t-n), \tag{67}$$

$$\beta(t) := T_1 y(t-1) + T_2 y(t-2) + \cdots + T_n y(t-n) \tag{68}$$

$$\hat{\overline{F}} := \left[\hat{\overline{f}}_1, \hat{\overline{f}}_2, \ldots, \hat{\overline{f}}_{\mathrm{n}}\right]^T \in \mathbb{R}^n \tag{69}$$

$$\hat{\theta} := \left[\hat{\overline{F}}^T, \hat{\overline{G}}^T, \hat{d}, \hat{J}^T\right]^T \in \mathbb{R}^{n_s}, n_s = 2n + 1 + n_J \tag{70}$$

$$xp(t+1) = \hat{\overline{F}} xp(t) + \hat{\overline{G}} u(t) + Ty(t) \tag{71}$$

$$P = \hat{\overline{F}} * P * \hat{\overline{F}}^T + Q - \hat{T} * \hat{S}^T \tag{72}$$

$$K = (P * c)/c * P * c^T + \hat{R} \tag{73}$$

$$x(t+1) = xp(t+1) + +K\widehat{v^*}(t) \tag{74}$$

$$P = (eye(n) - Kc) * P \tag{75}$$

$$\widehat{v^*}(t) = y(t) - H\hat{x}(t) - \hat{d}(t)u(t), \tag{76}$$

$$\hat{v}(t) = \widehat{v^*}(t) - \hat{J}_1\hat{v}(t-1) - \hat{J}_2\hat{v}(t-2) \tag{77}$$

$$\hat{w}(t) = \hat{x}(t+1) - \hat{\overline{F}}(t)\hat{x}(t) - \hat{\overline{G}}(t)u(t) - T\left(cx(t) + \hat{d}\, u(t)\right) \tag{78}$$

$$\hat{\overline{w}}(t) = \hat{w}(t) - T\widehat{v^*}(t) \tag{79}$$

$$\hat{x}_1(t) = x(1,t). \tag{80}$$

Now, our target system is the system described by Eq. (1)–(2)

$$x(t+1) = Fx(t) + Gu(t) + w(t)$$
$$y(t) = Hx(t) + du(t) + v^*(t)$$

$$v^*(t) := \left(1 + J_1 q^{-1} + J_2 q^{-2} + \cdots + J_{n_J} q^{-n_J}\right) v(t)$$

We substitute

$$\hat{F} = \hat{\overline{F}} + \hat{T}H$$

$$\hat{G} = \left(\hat{\overline{G}} + \hat{T}d\right)$$

$$S = \rho_{w,v}\sqrt{R}\sqrt{Q},$$

$$\overline{Q} = Q - TS^T$$

$$\hat{F} := \begin{bmatrix} -\hat{f}_1 & 1 & 0 & \cdots & 0 \\ -\hat{f}_2 & 0 & 1 & \cdots & 0 \\ \vdots & \vdots & \vdots & \ddots & \vdots \\ -\hat{f}_{n-1} & 0 & 0 & \cdots & 1 \\ -\hat{f}_n & 0 & 0 & \cdots & 0 \end{bmatrix}, \widehat{G} := \begin{bmatrix} \hat{g}_1 \\ \hat{g}_2 \\ \vdots \\ \hat{g}_{n-1} \\ \hat{g}_n \end{bmatrix}.$$

Finally, we extract the system parameter estimates as:

$$\theta_{estimated} := \left[\hat{F}^T, \hat{G}^T, \hat{d}, \hat{J}^T\right]^T \in \mathbb{R}^{n_s}, n_s = 2n + 1 + n_J,$$

**Remark 4:** The KF-CN-RGELS algorithm demonstrates the utility of using correlation coefficients to obtain state estimates and improve the accuracy of parameter estimation. This innovative approach is pivotal in achieving accurate system estimation in cases where parameters and states are not entirely known.

## 7. The impact of negative correlation coefficients $\rho_{wv}$ on state estimation accuracy

In this section, we address the impact of negative correlation coefficients on estimation accuracy. We have studied this effect by examining its influence on two fundamental factors: observation accuracy and process model accuracy.

### 7.1 Observation accuracy

To commence our exploration of observation accuracy, let us examine the model and measurement equations under the assumptions $F = I, H = 1$, and $u(t) = 0$ in Eqs. (31) and (32). This examination leads us to the following relationship:

$$y(t) = y(t-1) + v^*(t) + \overline{w}(t-1) + v^*(t-1) \tag{81}$$

Considering the variance of $y(t)$, we can express it as:

$$\begin{aligned} \text{var}[y(t)] &= \text{var}[y(t-1)] + \text{var}[v^*(t)] \\ &+2\text{cov}[y(t-1), v^*(t)] + \text{var}[\overline{w}(t-1)] \\ &+\text{var}[v^*(t-1)] \\ &-2\text{cov}[\overline{w}(t-1), v^*(t-1)], \end{aligned} \tag{82}$$

From (82), The measurement covariance $P(y(t))$ in term of the noise covariances $\overline{Q}(t), R(t),$ and $S(t)$ can be equivalently written as

$$P(y(t)) = P(y(t-1)) + \overline{Q}(t-1) + R(t-1) - 2S(t-1) \tag{83}$$

For uncorrelated noises, i.e., $S(t) = 0$, the equation becomes:

$$P_0(y(t)) = P(y(t-1)) + \overline{Q}(t-1) + R(t-1) \tag{84}$$

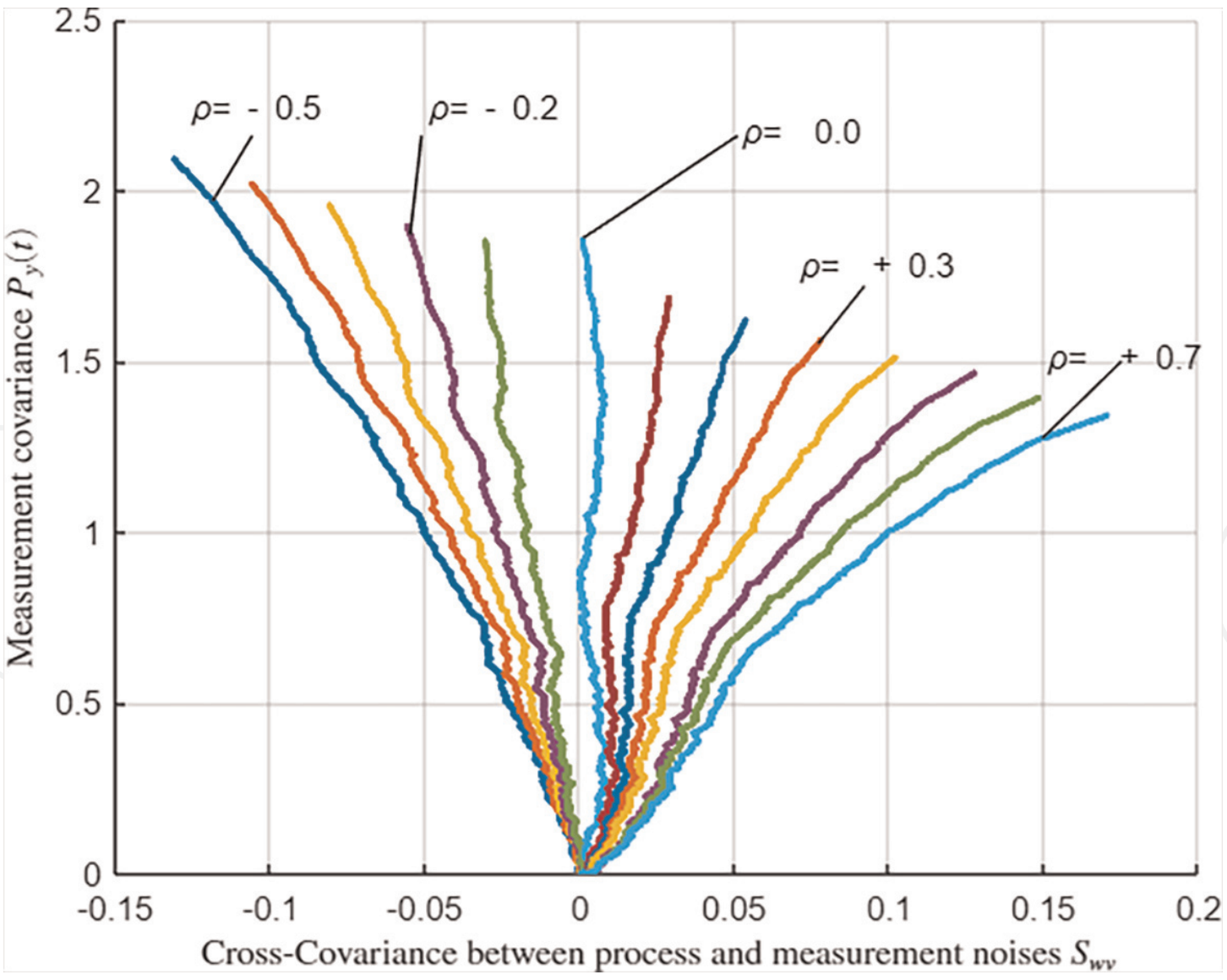

**Figure 1.**
*Noise cross-covariance, S versus measurement covariance $P(y(t))$ for different values of correlation coefficient $\rho_{w,v}$.*

According to (83) and (84), if $S(t) > 0$, then $P(y(t)) < P_0(y(t))$ indicates that the observation at time $t$ becomes more accurate when a positive correlation coefficient exists at time $t-1$. Conversely, if $S(t) < 0$ the observation at time $t$ becomes less accurate. It is worth noting that $\rho_{w,v} = S/\sqrt{Q}\sqrt{R}$. **Figure 1** below shows the measurement covariance $P(y(t))$ versus positive and negative values of the cross-covariance between process and measurement noises, $S(t)$. The data obtained correspond to illustrative example 1.

### 7.2 Process model accuracy

Turning to process model accuracy, we aim to derive the cross-covariance between the state and the process noise, defined as:

$$P(\hat{w}(k|k), \hat{x}(k|k)) \quad = E\left[\left(w(k) - \hat{w}(k|K)\right)(x(k) - \hat{x}(k|k))^T\right] \tag{85}$$

Before we delve into the derivation, we rely on a lemma known as the Conditional Gaussian Distribution Lemma presented in Ref. [36], which proves to be invaluable.

**Lemma 1:** Suppose a pair of vectors $Y$ and $X$ are jointly Gaussian with a mean vector $m_\Gamma$ and a covariance matrix $P_{\Gamma\Gamma}$, then $X$ is conditionally Gaussian on $Y$ with a conditional mean vector $m_{X|Y}$ and a conditional covariance matrix $P_{X|Y}$.

With this lemma, we can set:

$\Gamma = \begin{bmatrix} Y \\ X \end{bmatrix} = \begin{bmatrix} y(k) \\ w(k) \end{bmatrix} = \begin{bmatrix} Hx(k) + v(k) \\ w(k) \end{bmatrix}$, these yields

$m_\Gamma = \begin{bmatrix} \hat{y}(k|k-1) \\ \hat{w}(k|k-1) \end{bmatrix} = \begin{bmatrix} H\hat{x}(k|k-1) \\ 0 \end{bmatrix}$ And $P_{\Gamma\Gamma} = \begin{bmatrix} P_{y(k)y(k)} & P_{y(k)w(k)} \\ P_{w(k)y(k)} & P_{w(k)w(k)} \end{bmatrix}$

Where $P_{y(k)y(k)}$ is the measurement prediction covariance, $HP_{\hat{x}_{k|k-1}\hat{x}_{k|k-1}}H^T$, and $P_{w(k)y(k)}$ is the cross-covariance between process and measurement noise, $S$.

Hence, we can express:

$$P_{\Gamma\Gamma} = \begin{bmatrix} HP_{\hat{x}_{k|k-1}\hat{x}_{k|k-1}}H^T & S^T \\ S & Q \end{bmatrix}$$

Leveraging Lemma 1, we derive the conditional mean and covariance of the process noise.

$$\hat{w}(k|k) = \hat{w}(k|k-1) + S\left(P_{\hat{x}_{k|k-1}\hat{x}_{k|k-1}}H^T\right)^{-1}(y(k) - \hat{y}(k|k-1)) \tag{86}$$

And

$$P\left(\hat{W}(k|k) = Q - S\left(P_{\hat{x}_{k|k-1}\hat{x}_{k|k-1}}H^T\right)^{-1}S^T. \tag{87}$$

By using Eqs. (5) and (86) with (85), we can prove:

$$P(\hat{w}(k|k), \hat{x}(k|k)) = -S\left(H\,P_{\hat{x}_{k|k-1}\hat{x}_{k|k-1}}H^T\right)^{-1}HP_{\hat{x}_{k|k-1}\hat{x}_{k|k-1}} \tag{88}$$

The results presented in Eq. (88) highlight the role of the cross-covariance between the state and process noise in the context of the correlation coefficient. The correlation coefficient can be both positive or negative values and significantly impacts this cross-covariance. When the correlation coefficient is positive, increasing its value reduces the cross-covariance between the state and process noise. Conversely, when the correlation coefficient is negative, increasing its value elevates the cross-covariance between the state and process noise. **Table 1** and **Figure 2** further illustrate this outcome. The data obtained corresponds to illustrative example 1.

| $\rho_{w,v}$ | −0.7 | −0.5 | −0.3 | 0 | 0.3 | 0.5 | 0.7 |
|---|---|---|---|---|---|---|---|
| Tr $[P(\hat{w}(k)|k), \hat{x}(k|k)]$ | 0.0166 | 0.0097 | 0.0052 | −0.0001 | −0.0038 | −0.0059 | −0.0076 |

**Table 1.**
*Correlation coefficient $\rho_{w,v}$ versus trace of cross-covariance $P(\hat{w}(k|k), \hat{x}(k|k))$ between state and process noise.*

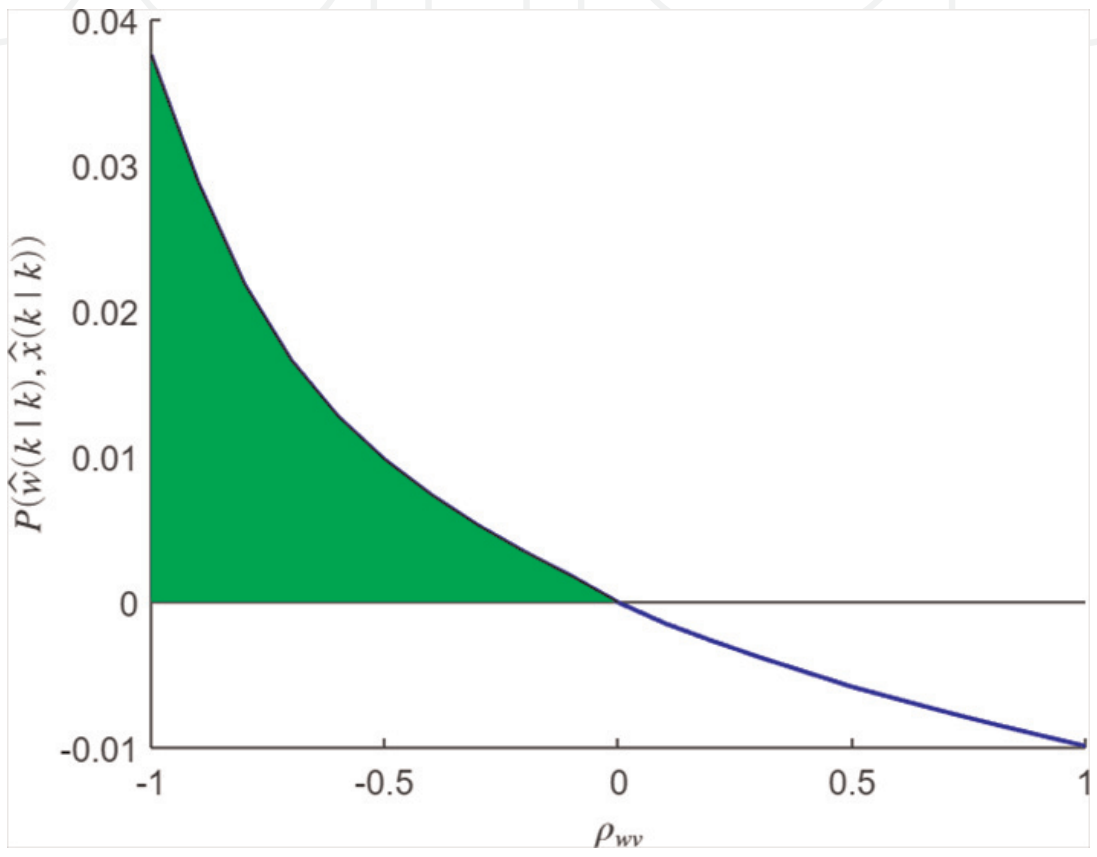

**Figure 2.**
*Correlation coefficient $\rho_{w,v}$ versus trace of cross-covariance $P(\hat{w}(k|k), \hat{x}(k|k))$ between state and process noise.*

## 8. Illustrative examples

Example 1. Consider the observer canonical state-space description.

$$x(t+1) = Fx(t) + Gu(t) + w(t),$$
$$y(t) = Hx(t) + du(t) + J(q)v(t).$$

$$F = \begin{bmatrix} -f_1 & 1 \\ -f_2 & 0 \end{bmatrix} = \begin{bmatrix} 0.05 & 1 \\ 0.35 & 0 \end{bmatrix}, G = \begin{bmatrix} g_1 \\ g_2 \end{bmatrix} = \begin{bmatrix} 2.0 \\ 3.0 \end{bmatrix}, H = [1, \quad 0], d = 1.30, w(t) = \begin{bmatrix} w_1(t) \\ w_2(t) \end{bmatrix}$$

$$J(q) = 1 + J_1 q^{-1} + J_2 q^{-2} = 1 + 0.0505q^{-1} + 0.0139q^{-2}.$$

The parameter vector to be identified is given by:

$$\begin{aligned} \theta &= [f_1, f_2, g_1, g_2, d, J_1, J_2]^T, \\ &= [-0.05, -0.35, 2.0, 3.0, 1.30, 0.0505, 0.0139]^T. \end{aligned}$$

- For reliable models, choosing system parameters must guarantee stability (avoiding unbounded oscillations), controllability (being able to reach any desired state), and observability (knowing the internal state from measurable outputs).

- In the simulation, the input $\{u(t)\}$ is a pseudo-random binary sequence generated by the MATLAB function $u =$ idinput ([65535, 1, 1], prbs' ', [0, 1], [−0.8,1]), $w_1(t)$ and $w_2(t)$ are random noise sequences with zero mean and variance $\sigma_{w_1}^2 = 0.07^2$, and $\sigma_{w_2}^2 = 0.01^2$ respectively. $v(t)$ is a random noise sequence with zero mean and variance $\sigma_v^2 = 0.8^2$. Set the data length $L = 5000$ and choose different values of the correlation coefficient $\rho_{w,v}$ in the range [0–1]. Generate system parameter and state estimates by applying the KF-CNRGELS algorithm with different values of correlation coefficient, $\rho_{w,v}$ and examine the correlation coefficient effect.

- To facilitate simulation, the generation of correlated process and measurement noises involves creating a correlation matrix. This matrix is then applied to eigen decomposition, giving rise to a correlating filter that captures the interdependencies within the noise components.

  The parameters estimates and errors $\delta_\theta = \|\hat{\theta} - \theta\| / \|\theta\|$ at $\rho_{w,v} = 0.5, 0.6$, and 0.9 are summarized in **Table 2**.

- In **Figure 3**, we vary the correlation coefficient $\rho_{w,v}$ and plot the resulting parameter estimation errors. The plot shows three curves corresponding to correlation coefficients of 0, 0.5, and 0.9.

- The root-mean-squared error of the state estimate $x_1$ for $\rho_{w,v}$ in the range $-$ 1.0 to +1.0 is shown in **Figure 4**. **Figure 5** shows the relationship between the parameter estimation error and various values of the correlation coefficient $\rho_{w,v}$ positive and negative values. **Figure 6** shows the parameter estimation error of the proposed algorithm compared to the standard Kalman filter SKF and Augmented-State Kalman filter algorithms for $\rho_{w,v} = 0.6$. **Tables 3** and **4** illustrate the results of the comparison between the KF-CN-GELS and AUG-KF for $\rho_{w,v} = 0.0$, the KF-CN-GELS and AUG-KF for $\rho_{w,v} = 0.4$ and the KF-CN-GELS and AUG-KF for $\rho_{w,v} = 0.8$. The noise estimates $\hat{v}(t), \hat{w}_1(t)$ and $\hat{w}_2(t)$ for KF-CN-GELS with $\rho_{w,v} = 0.7$ is shown in **Figure 7**. The state estimates of $x_1(t)$and $x_2(t)$ and errors for $\rho_{w,v} = 0.7$ are depicted in **Figure 8**. The collected input and output data are shown in **Figure 9**.

| $\rho_{w,v}$ | $t$ | $f_1$ | $f_2$ | $g_1$ | $g_2$ | $d$ | $J_1$ | $J_2$ | $\delta_\theta\%$ |
|---|---|---|---|---|---|---|---|---|---|
| 0.5 | 5000 | −0.0402 | −0.3557 | 2.0117 | 3.0165 | 1.3191 | 0.0580 | 0.0398 | 1.0491 |
| 0.6 | 5000 | −0.0396 | −0.3557 | 2.0110 | 3.0173 | 1.3185 | 0.0587 | 0.0345 | 0.9696 |
| 0.9 | 5000 | 0.0416 | −0.3545 | 2.0098 | 3.0148 | 1.3120 | 0.0470 | 0.0201 | 0.6354 |
| True values | | 0.0500 | −0.3500 | 2.0000 | 3.0000 | 1.3000 | 0.0505 | 0.0139 | |

**Table 2.**
*The KF-CN-RGELS estimates and errors* $(\rho_{w,v} = 0.5, 0.6,$ *and* $0.9$ *for* $(R_v = 0.8, Q_w = [0.07, 0.01]I_2)$.

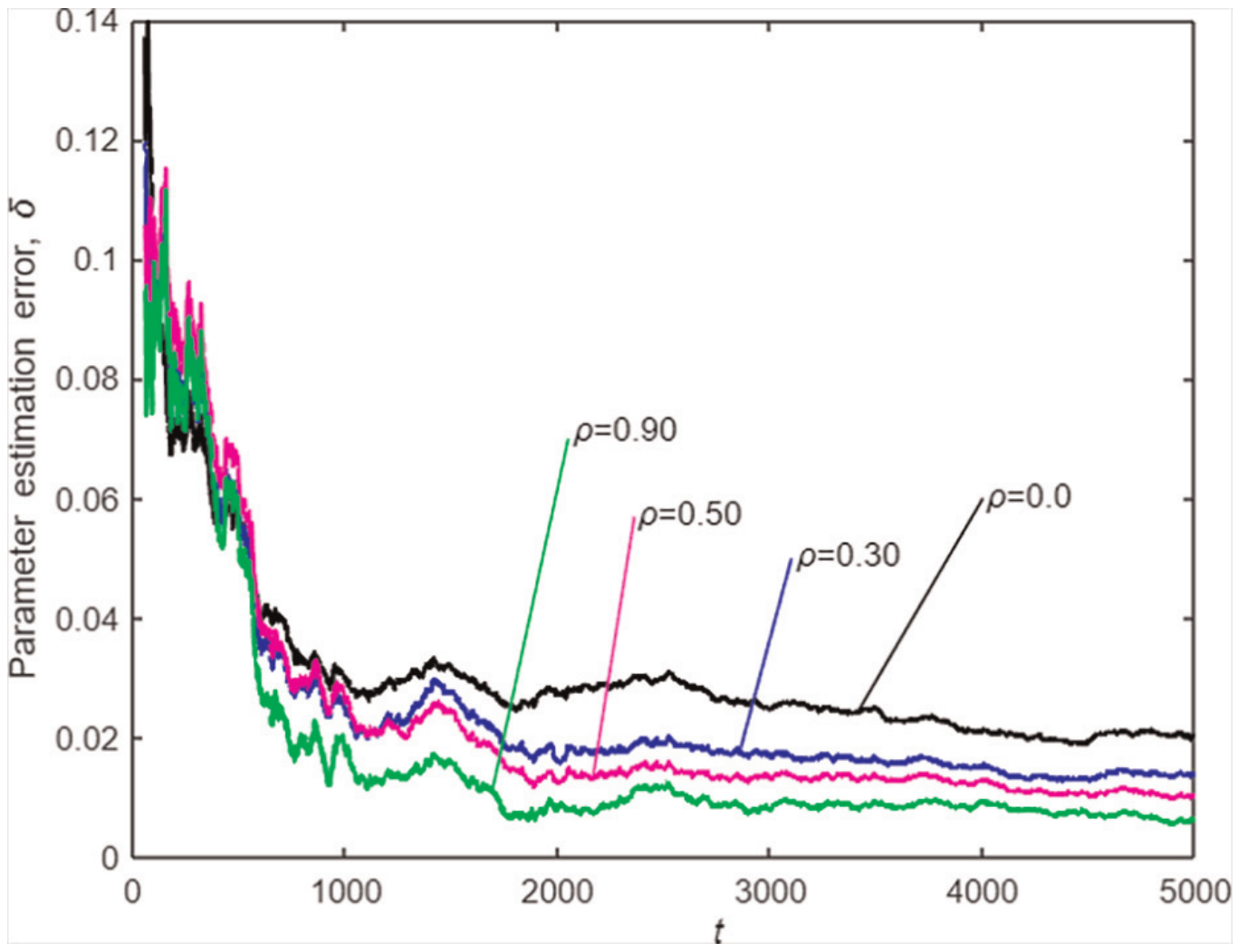

**Figure 3.**
*The KF-CN-RGELS parameter estimation error* $\delta_\theta$. *Against* $t(\rho_{w,v} = 0, 0.3, 0.5,$ *and* $0.9$, *and* $R_v = 0.8, Q_w = [0.07, 0.01]I_2.)$.

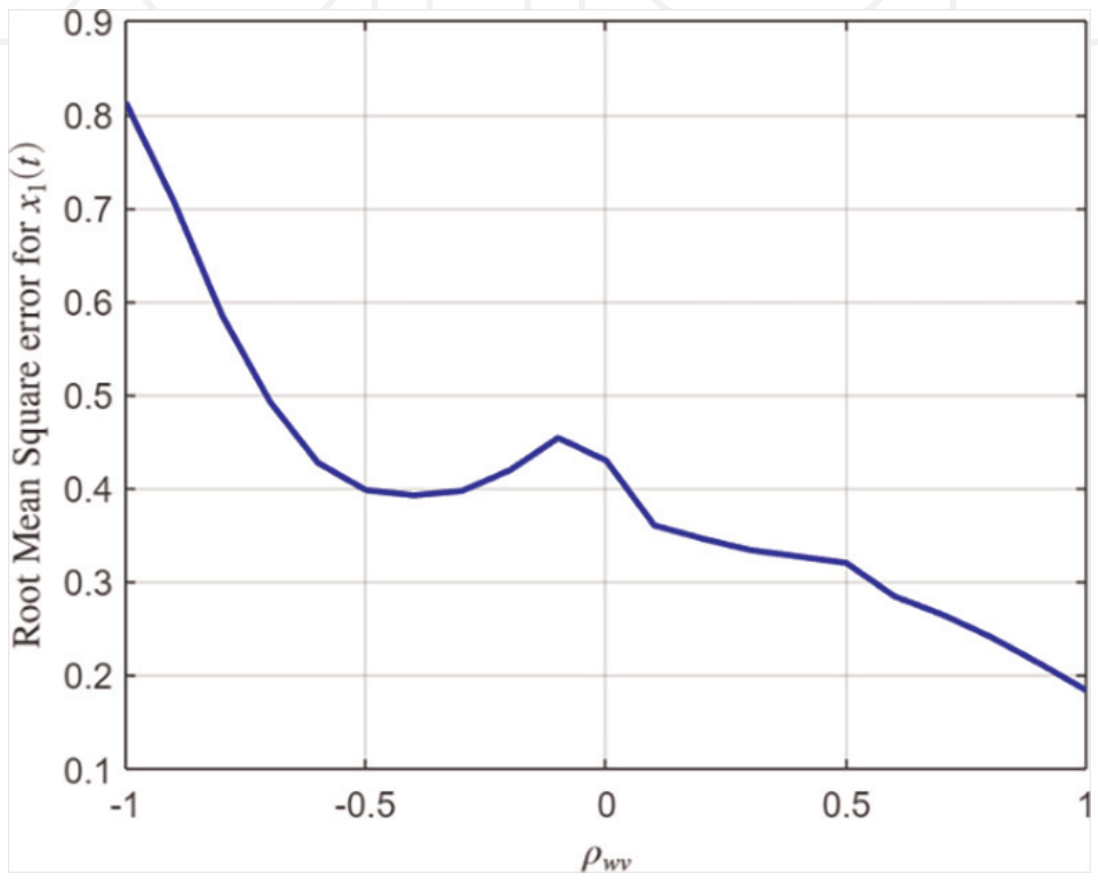

**Figure 4.**
*Root mean square error (RMSE) for* $x_1(t)$ *versus different values of correlation coefficient* $\rho_{w,v}$.

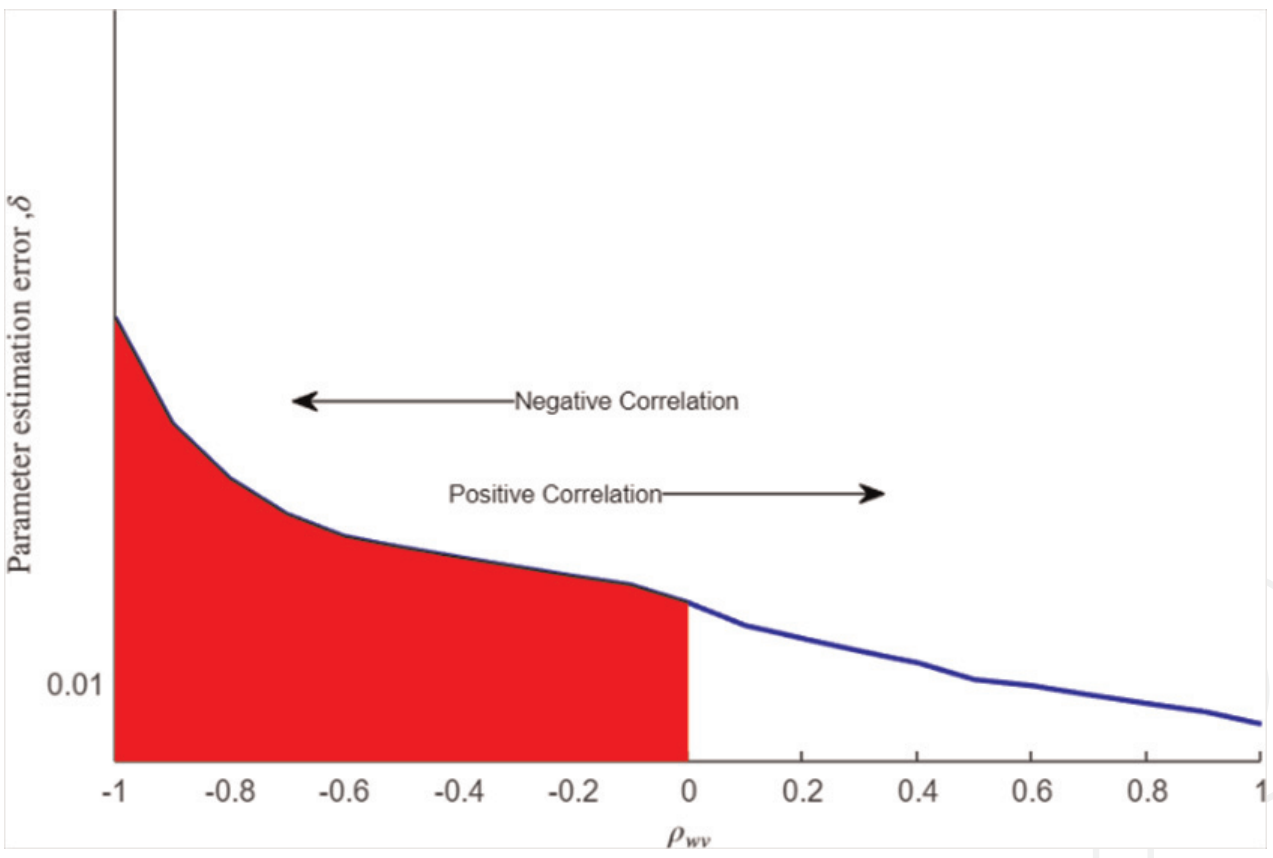

**Figure 5.**
*The parameter estimation error $\delta_\theta$ against $\rho_{w,v} = -1$ to $+1$, $(R_v = 0.15, Q_w = [0.07,0.01]I_2)$.*

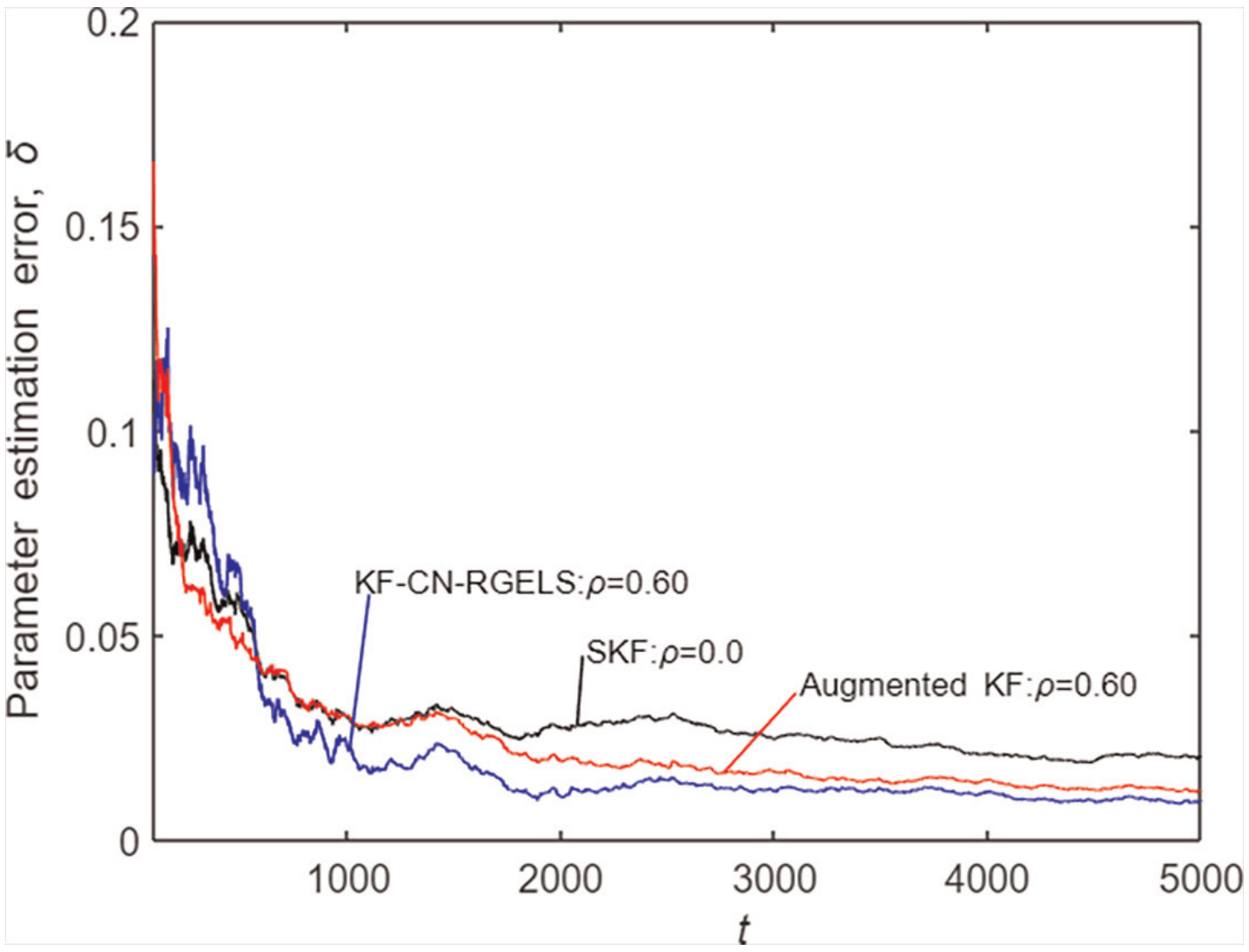

**Figure 6.**
*The KF-CN-RGELS estimation error $\delta_\theta$ against t compared to SKF and augmented KF algorithms $(\rho_{w,v} = 0.60$ for $R_v = 0.8, Q_w = [0.07,0.01]I_2)$.*

| $\rho_{w,v} = 0$ | $t$ | $f_1$ | $f_2$ | $g_1$ | $g_2$ | $d$ | $J_1$ | $J_2$ | $\delta_\theta\%$ |
|---|---|---|---|---|---|---|---|---|---|
| SKF | 5000 | −0.0521 | −0.3461 | 2.0186 | 3.0047 | 1.3108 | 0.0967 | 0.0732 | 2.0386 |
| AUG-KF | 5000 | −0.0522 | −0.3461 | 2.0186 | 3.0047 | 1.3108 | 0.0967 | 0.0732 | 2.0389 |
| KF-CN-RGELS | 5000 | −0.0524 | −0.3464 | 2.0190 | 3.0052 | 1.3108 | 0.0958 | 0.0727 | 2.0177 |
| True values | | 0.0500 | −0.3500 | 2.0000 | 3.0000 | 1.3000 | 0.0505 | 0.0139 | |

**Table 3.**
*The parameter estimates and errors (KF-CN-RGELS, SKF, and AUG-KF) for $\rho_{wv} = 0.0$ $(R_v = 0.8, Q_w = [0.07,0.01]I_2)$.*

| Algorithms | KF-CN-RGELS ($\rho_{wv} = 0.4$) | AUG-KF ($\rho_{wv} = 0.4$) | KF-CN-RGELS ($\rho_{wv} = 0.6$) | AUG-KF ($\rho_{wv} = 0.6$) | KF-CN-RGELS ($\rho_{wv} = 0.8$) | AUG-KF ($\rho_{wv} = 0.8$) |
|---|---|---|---|---|---|---|
| $f_1 = -0.0500$ | −0.0388 | −0.0413 | −0.0396 | −0.0446 | −0.0408 | −0.0515 |
| $f_2 = -0.3500$ | −0.3553 | −0.3537 | −0.3557 | −0.3511 | −0.3549 | −0.3456 |
| $g_1 = 2.0000$ | 2.0098 | 2.0103 | 2.0110 | 2.0091 | 2.0101 | 2.0090 |
| $g_2 = 3.0000$ | 3.0164 | 3.0132 | 3.0173 | 3.0080 | 3.0158 | 2.9971 |
| $d = 1.3000$ | 1.3221 | 1.3218 | 1.3185 | 1.3193 | 1.3146 | 1.3170 |
| $J_1 = 0.0505$ | 0.0698 | 0.0740 | 0.0587 | 0.0570 | 0.0500 | 0.2615 |
| $J_2 = 0.0139$ | 0.0452 | 0.0580 | 0.0345 | 0.0537 | 0.0256 | 0.1422 |
| $\delta_\theta(\%)$ | 1.2622 | 1.5008 | 0.9696 | 1.2109 | 0.7413 | 6.4351 |

**Table 4.**
*The parameter estimates and errors (KF-CN-RGELS and AUG-KF) for* $(R_v = 0.8, Q_w = [0.07, 0.01]I_2)$.

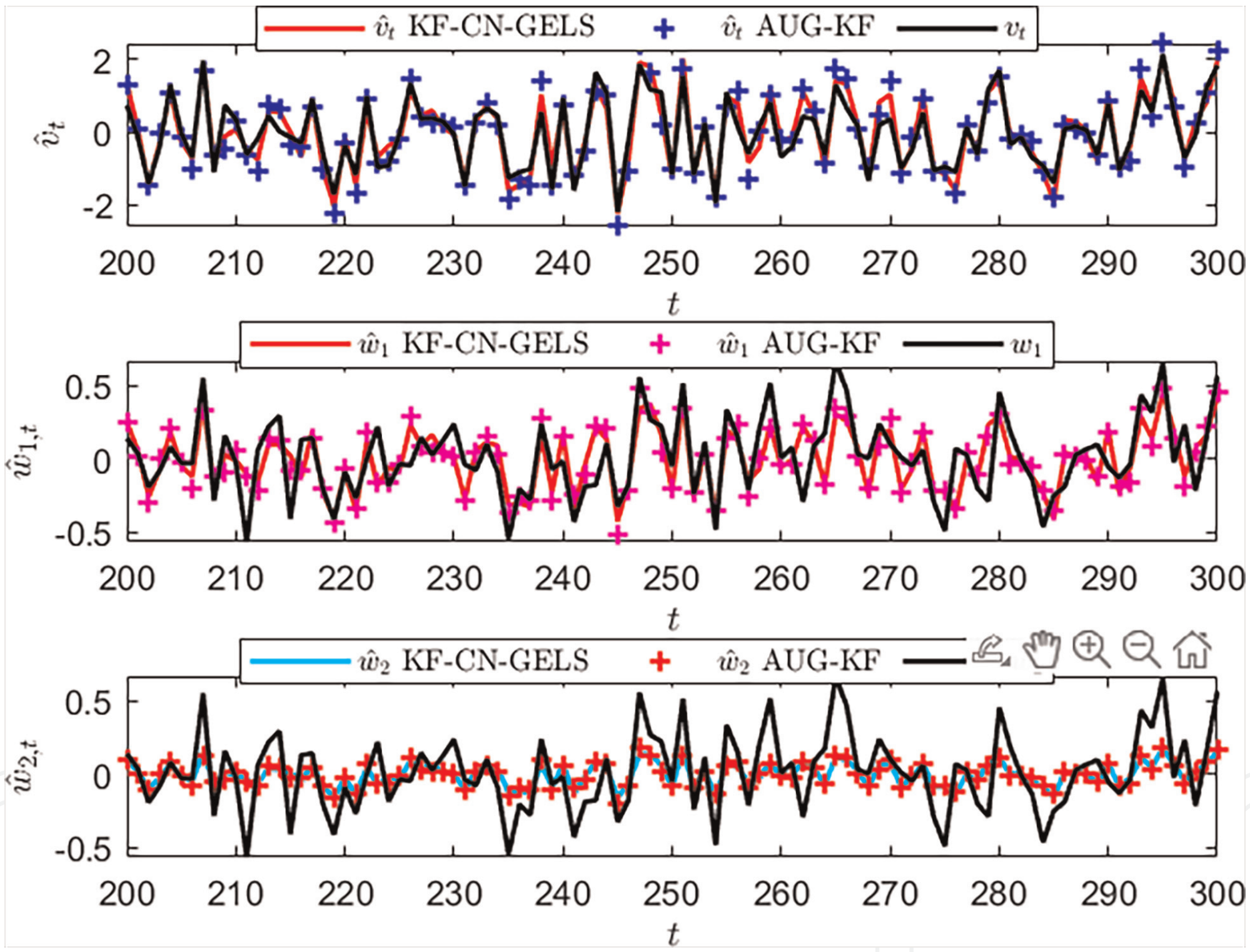

**Figure 7.**
*The estimates of* $w_2(t), w_2(t)$, *and* $v(t)$ *for KF-CN-GELS and AUG-KF* $(\rho_{w,v} = 0.7$ *for* $R_v = 0.8, Q_w = [0.07, 0.01]I_2)$.

Looking at **Tables 2–4** and **Figures 3–9**, we can draw some conclusions from these tables and figures.

- At $\rho_{w,v} = 0$, there is no correlation between process noise and measurement noise. The filtering result of the standard Kalman filter is the same as the filtering result of the KF-CN-RGELS and the AUG-KF algorithms, as illustrated in **Table 3**.

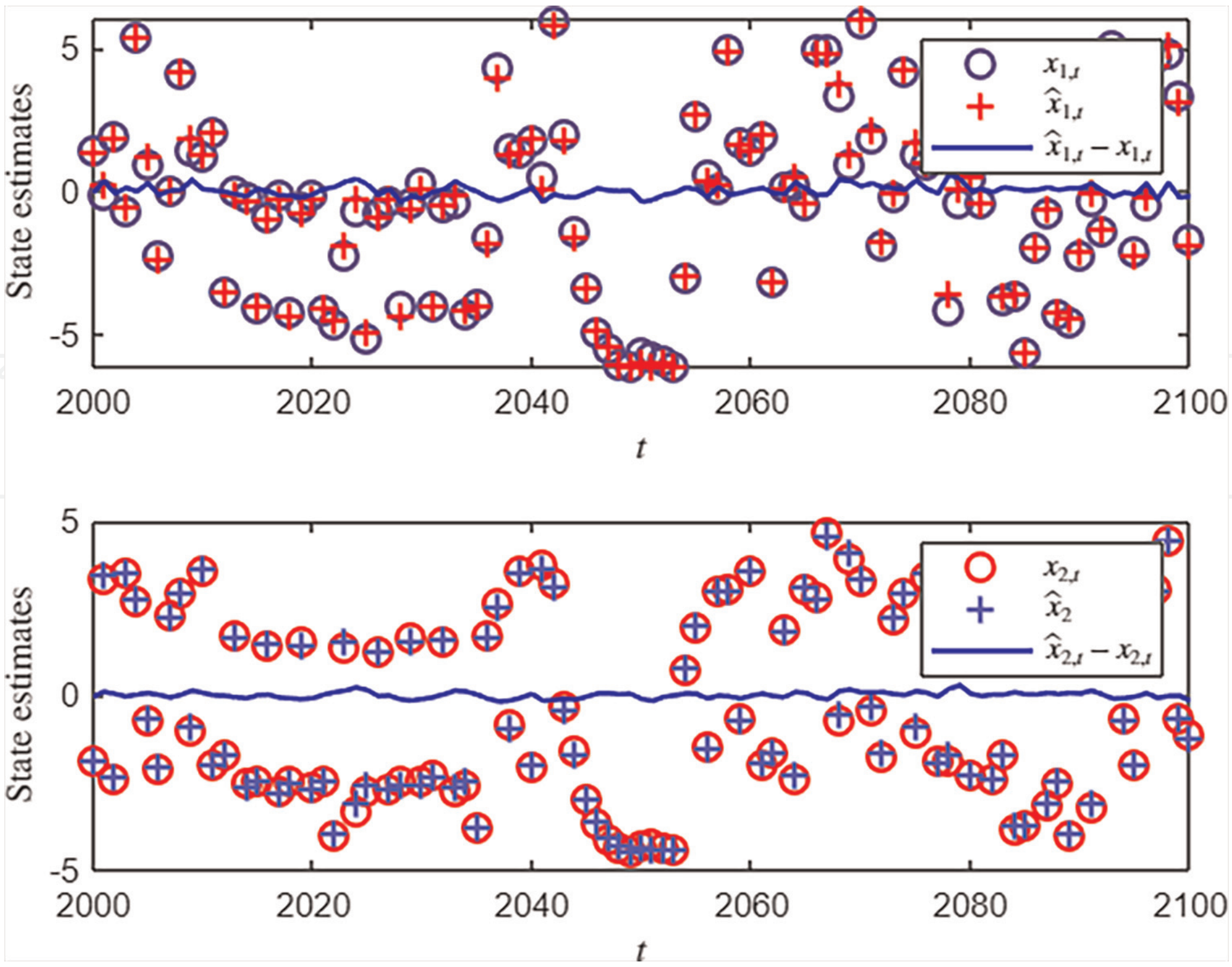

**Figure 8.**
*The true state, estimated state, and state estimation error used in* ***example 1***, $\rho_{w,v} = 0.70$.

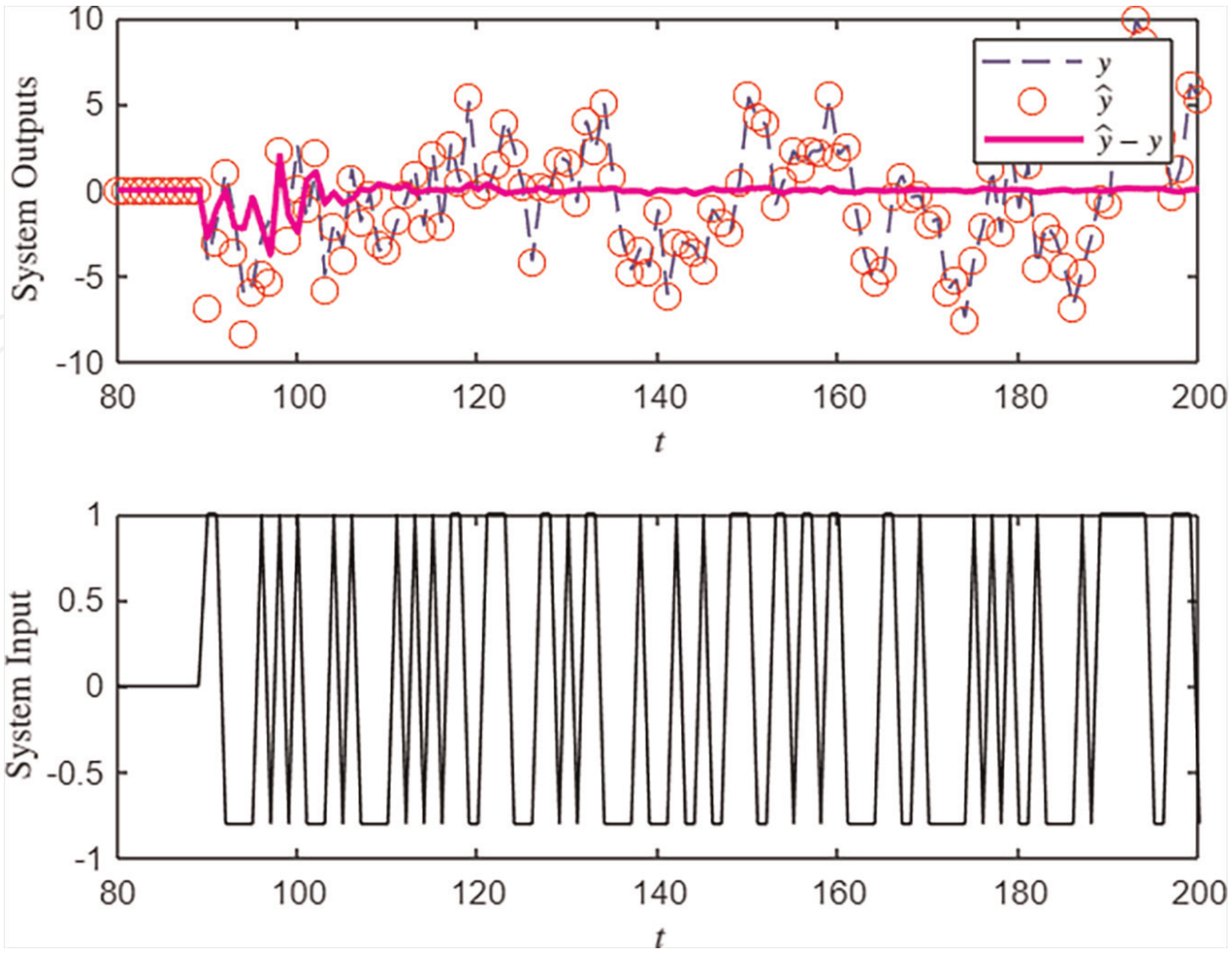

**Figure 9.**
*The input $u(t)$ and output $y(t)$ collected data used in* ***example 1*** $\rho_{w,v} = 0.70$.

- The best estimate of the state is obtained when the correlation coefficient $\rho_{w,v}$ between the process noise and the measurement noise increases. This is reflected in the improved accuracy of the parameter estimates presented by the KF-CN-RGELS algorithm. See **Figure 4** and **Table 2**.

- Clearly, the KF-CN-RGELS algorithm yields more accurate parameter estimates, since the correlation coefficient $\rho_{w,v}$ increases in the positive direction, and the mean-square error of the estimated states decreases. See **Figures 4** and **5** and **Table 2**.

- We find that the KF-CN-RGELS algorithm provides better parameter estimation accuracy than the SKF and AUG-KF algorithms under the same conditions, which makes this algorithm efficient and robust. See **Figure 6** and **Table 4**.

- The KF-CN-GELS algorithm provides better estimates of the process and measurement noises. $w_1(t)$, $w_2(t)$, and $v(t)$ than the AUG-KF algorithms under the same conditions. **Figure 7** shows these results.

Example 2. Consider the state-space model for a simple two-tank system, where $u(t)$ is the inlet water flow, $x_1(t)$ and $x_2(t)$ are the water levels of two tank, $y(t)$ is measurement of water level of tank 1 as shown in **Figure 10**. The state-space model of this system includes process and measurement noises:

$$\begin{bmatrix} x_1(t+1) \\ x_2(t+1) \end{bmatrix} = \begin{bmatrix} 0.11 & 1 \\ 0.15 & 0 \end{bmatrix} \begin{bmatrix} x_1(t) \\ x_2(t) \end{bmatrix} + \begin{bmatrix} 1.9 \\ 1.6 \end{bmatrix} u(t) + \begin{bmatrix} \omega_1(t) \\ \omega_2(t) \end{bmatrix}$$

$$y(t) = \begin{bmatrix} 1 & 0 \end{bmatrix} \begin{bmatrix} x_1(t) \\ x_2(t) \end{bmatrix} + 1.9u(t) + \left[1 + 0.1069q^{-1} - 0.0143q^{-2}\right]v(t)$$

The parameter vector to be identified is given by:

$$\begin{aligned} \theta &= [f_1, f_2, g_1, g_2, d, J_1, J_2]^T \\ &= [-0.11, -0.15, 1.90, 1.60, 1.90, 0.1069, -0.0143]^T \end{aligned}$$

The $Q$ and $R$ matrices are assumed to be.

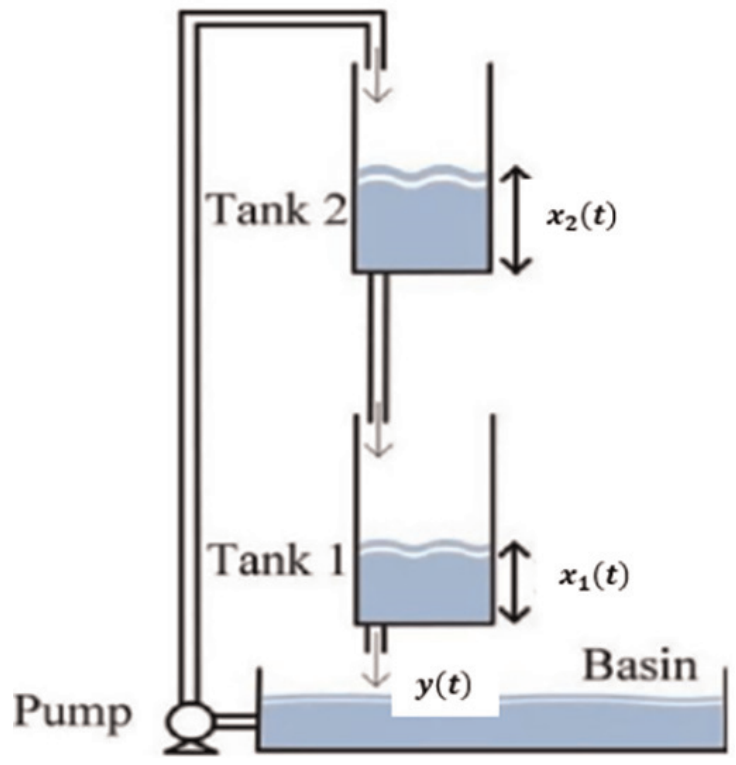

**Figure 10.**
*The two-tank schematic diagram used in example 2.*

$$Q = \begin{bmatrix} 0.60 & 0.00 \\ 0.00 & 0.40 \end{bmatrix}, R = [1.60], \text{respectively.}$$

In the case of correlated noises, we define the system noise matrix as

$$U = \begin{bmatrix} Q & S \\ S^T & R \end{bmatrix},$$

From the code for $\boldsymbol{\rho_{wv}} = \mathbf{0.95}, U = \begin{bmatrix} 0.5874 & 0.4813 & \mathbf{0.9798} \\ 0.4813 & 0.3947 & \mathbf{0.7999} \\ 0.9798 & 0.7999 & 1.7013 \end{bmatrix}$

The theoretical value of S is:

$S = \rho_{wv}\sqrt{Q}\sqrt{R} = 0.95 \times \sqrt{[0.6; 0.4]} \times \sqrt{1.6} = \begin{bmatrix} \mathbf{0.9308} \\ \mathbf{0.7600} \end{bmatrix}$, which is approximately $\begin{bmatrix} \boldsymbol{U(1,3)} \\ \boldsymbol{U(2,3)} \end{bmatrix}$.

During simulation, the input and output data that are collected from a pseudo-random binary sequence denoted by $u(t)$ and the water level measurement of tank1 $y(t) = x_1(t)$ are shown in **Figure 10**, $w_1(t)$ and $w_2(t)$ are white noise sequences with zero mean and variances $\sigma^2_{w_1} = 0.60^2$ and $\sigma^2_{w_2} = 0.40^2$ respectively. $v(t)$ is a white noise sequence with zero mean and variance $\sigma^2_v = 1.60^2$.Set the data length $L = 5000$. Apply the KF-CN-RGELS algorithm to identify this two-tank model. To test the performance of the algorithm, different values of correlation coefficient (positive and negative values) are used for the system, and the simulation results are displayed in **Tables 5–8** and **Figure 11**.

When choosing the correlation coefficient for the simulation, it is important to make sure that the noise matrix $U$ above is positive semidefinite, which means that all its eigenvalues are positive and the eigenvectors are orthogonal.

| $\rho_{w,v}$ | $t$ | $f_1$ | $f_2$ | $g_1$ | $g_2$ | $d$ | $J_1$ | $J_2$ | $\delta_\theta\%$ |
|---|---|---|---|---|---|---|---|---|---|
| 0.53 | 5000 | −0.1293 | −0.1322 | 1.8892 | 1.5373 | 1.9070 | 0.1536 | −0.0914 | 3.6241 |
| 0.75 | 5000 | −0.0960 | −0.1591 | 1.8915 | 1.6051 | 1.9030 | 0.1263 | −0.0990 | 2.8425 |
| True values | | −0.1100 | −0.1500 | 1.9000 | 1.6000 | 1.9000 | 0.1069 | −0.0143 | |

**Table 5.**
*The KF-CN-RGELS estimates and errors* $(\rho_{w,v} = 0.53$ *and* $0.75$ *for* $(R_v = 1.6, Q_w = [0.6, 0.4]I_2)$.

| $\rho_{w,v} = \mathbf{0}$ | $t$ | $f_1$ | $f_2$ | $g_1$ | $g_2$ | $d$ | $J_1$ | $J_2$ | $\delta_\theta\%$ |
|---|---|---|---|---|---|---|---|---|---|
| SKF | 5000 | −0.1213 | −0.1418 | 1.8904 | 1.5702 | 1.9004 | 0.3480 | 0.0832 | 8.3683 |
| AUG-KF | 5000 | −0.1216 | −0.1418 | 1.8903 | 1.5693 | 1.9003 | 0.3360 | 0.0721 | 7.8924 |
| KF-CN-RGELS | 5000 | −0.1221 | −0.1413 | 1.8901 | 1.5676 | 1.9003 | 0.3301 | 0.0607 | 7.6053 |
| True values | | −0.1100 | −0.1500 | 1.9000 | 1.6000 | 1.9000 | 0.1069 | −0.0143 | |

**Table 6.**
*The parameter estimates and errors (KF-CN-RGELS, SKF, and AUG-KF) for* $\rho_{wv} = 0.0$ $(R_v = 1.6, Q_w = [0.6, 0.4]I_2)$.

| Algorithms | KF-CN-RGELS ($\rho_{wv}=0.3$) | AUG-KF ($\rho_{wv}=0.3$) | KF-CN-RGELS ($\rho_{wv}=0.5$) | AUG-KF ($\rho_{wv}=0.5$) | KF-CN-RGELS ($\rho_{wv}=0.7$) | AUG-KF ($\rho_{wv}=0.7$) |
|---|---|---|---|---|---|---|
| $f_1 = -0.1100$ | −0.1487 | −0.1242 | −0.1336 | −0.0997 | −0.1019 | −0.0932 |
| $f_2 = -0.1500$ | −0.1181 | −0.1424 | −0.1286 | −0.1588 | −0.1546 | −0.1600 |
| $g_1 = 1.9000$ | 1.8897 | 1.9012 | 1.8888 | 1.8969 | 1.8914 | 1.8957 |
| $g_2 = 1.6000$ | 1.5047 | 1.5695 | 1.5286 | 1.6083 | 1.5934 | 1.6161 |
| $d = 1.9000$ | 1.9103 | 1.9097 | 1.9076 | 1.9062 | 1.9037 | 1.9057 |
| $J_1 = 0.1069$ | 0.2016 | 0.3822 | 0.1586 | 0.2400 | 0.1316 | 0.2045 |
| $J_2 = -0.0143$ | −0.0409 | 0.1490 | −0.0871 | 0.0201 | −0.1004 | −0.0001 |
| $\delta_\theta(\%)$ | 4.6767 | 10.2745 | 3.8111 | 4.4205 | 2.8971 | 3.2575 |

**Table 7.**
*The parameter estimates and errors (KF-CN-RGELS and AUG-KF) for* $(R_v = 1.6, Q_w = [0.6,0.4]I_2)$.

| Algorithms | $t$ | $f_1$ | $f_2$ | $g_1$ | $g_2$ | $d$ | $J_1$ | $J_2$ | $\delta_\theta\%$ |
|---|---|---|---|---|---|---|---|---|---|
| $KF-CN-RGELS_{\rho_{wv}=-0.5}$ | 5000 | −0.0241 | −0.1249 | 1.9996 | 1.8662 | 1.9901 | 0.4607 | 0.3575 | 19.1496 |
| $AUG-KF_{\rho_{wv}=-0.5}$ | 5000 | −0.1245 | −0.1432 | 1.8957 | 1.5831 | 1.8937 | 0.3581 | 0.0998 | 8.8369 |
| True values | | −0.1100 | −0.1500 | 1.9000 | 1.6000 | 1.9000 | 0.1069 | −0.0143 | |

**Table 8.**
*The parameter estimates and errors (KF-CN-RGELS and AUG-KF) for negative values of correlation coefficient* $\rho_{wv} = -0.5$ $(R_v = 1.6, Q_w = [0.6,0.4]I_2)$.

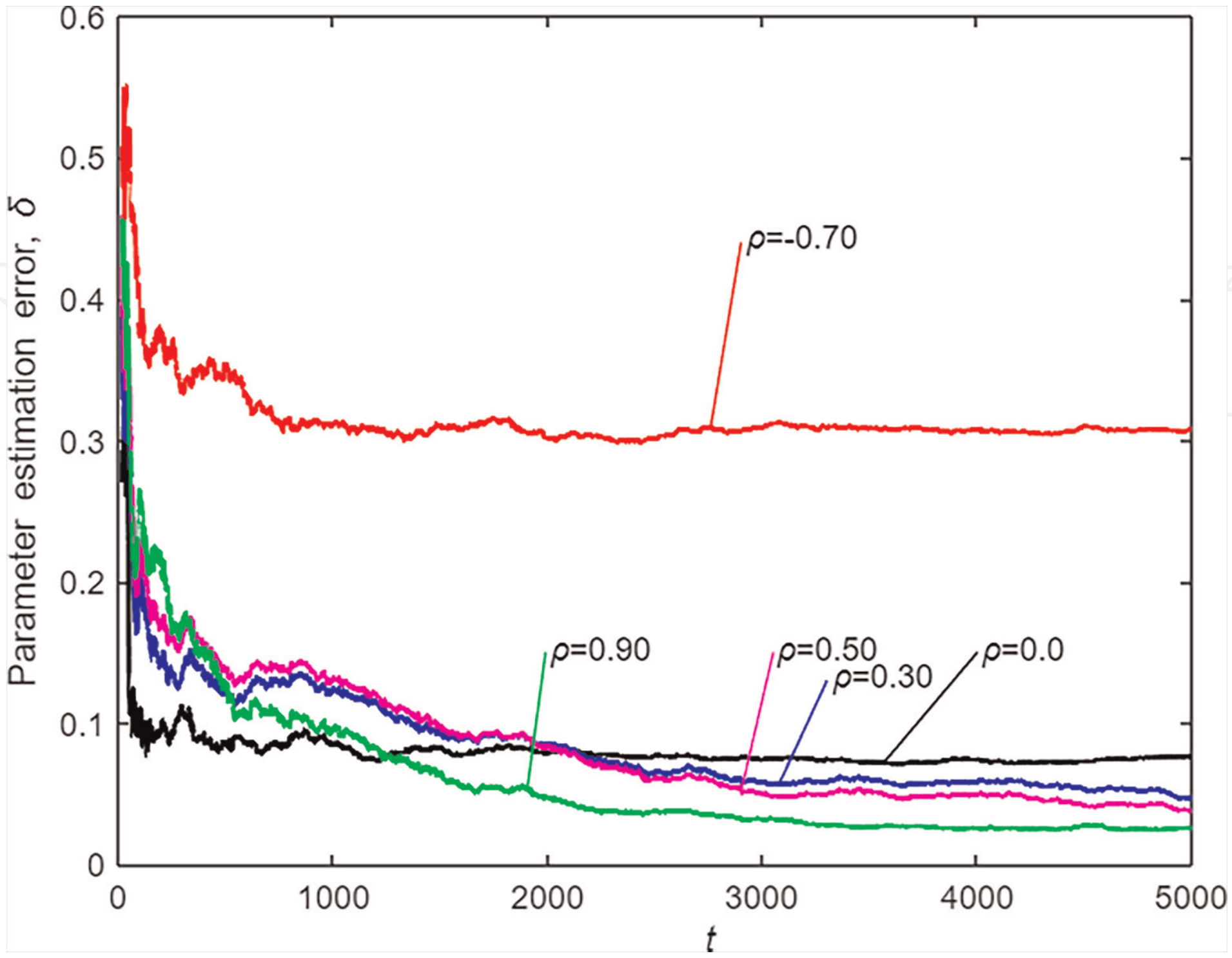

**Figure 11.**
*The KF-CN-RGELS parameter estimation error* $\delta_\theta$.

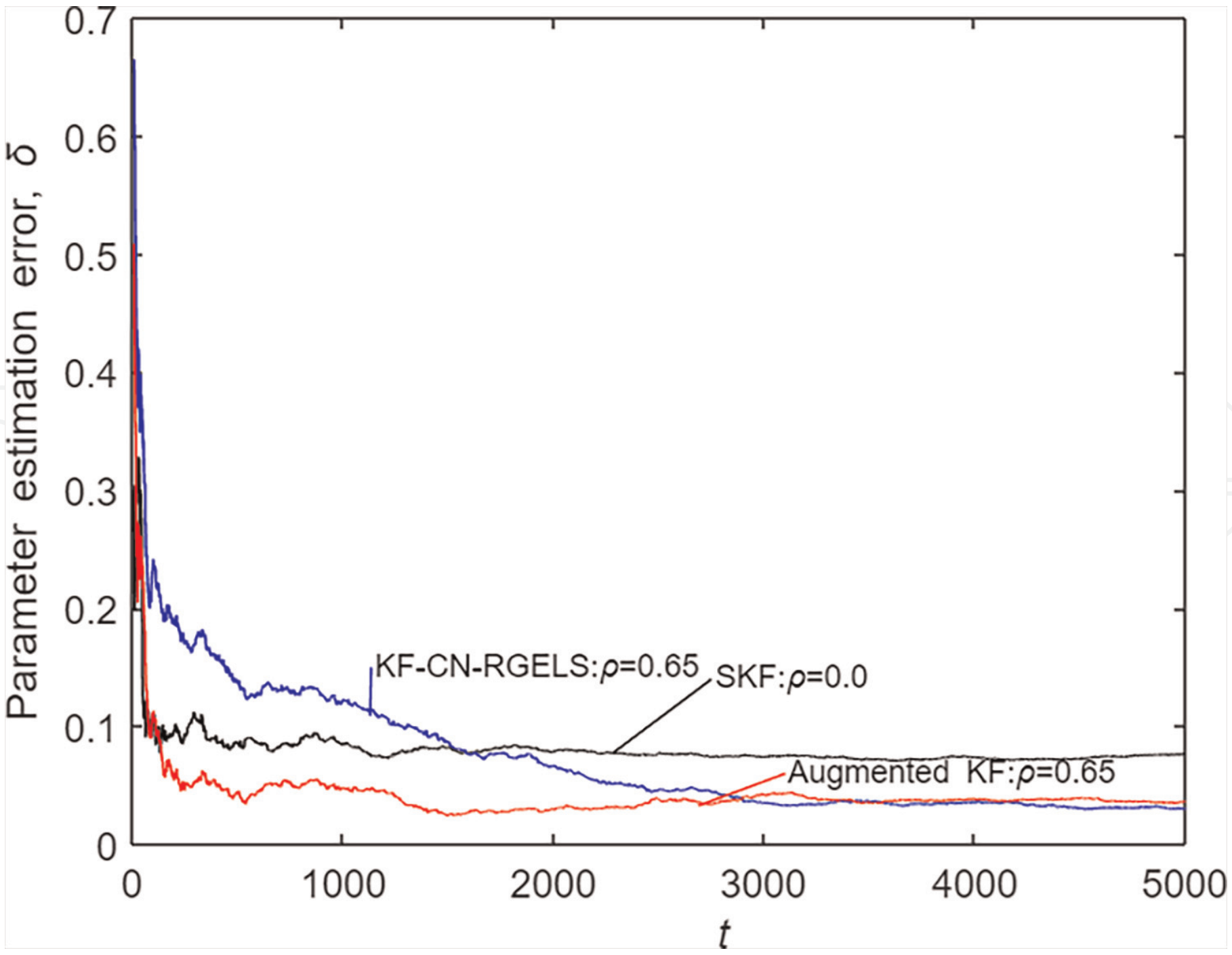

**Figure 12.**
*The KF-CN-RGELS estimation error $\delta_\theta$ against t compared to SKF.*

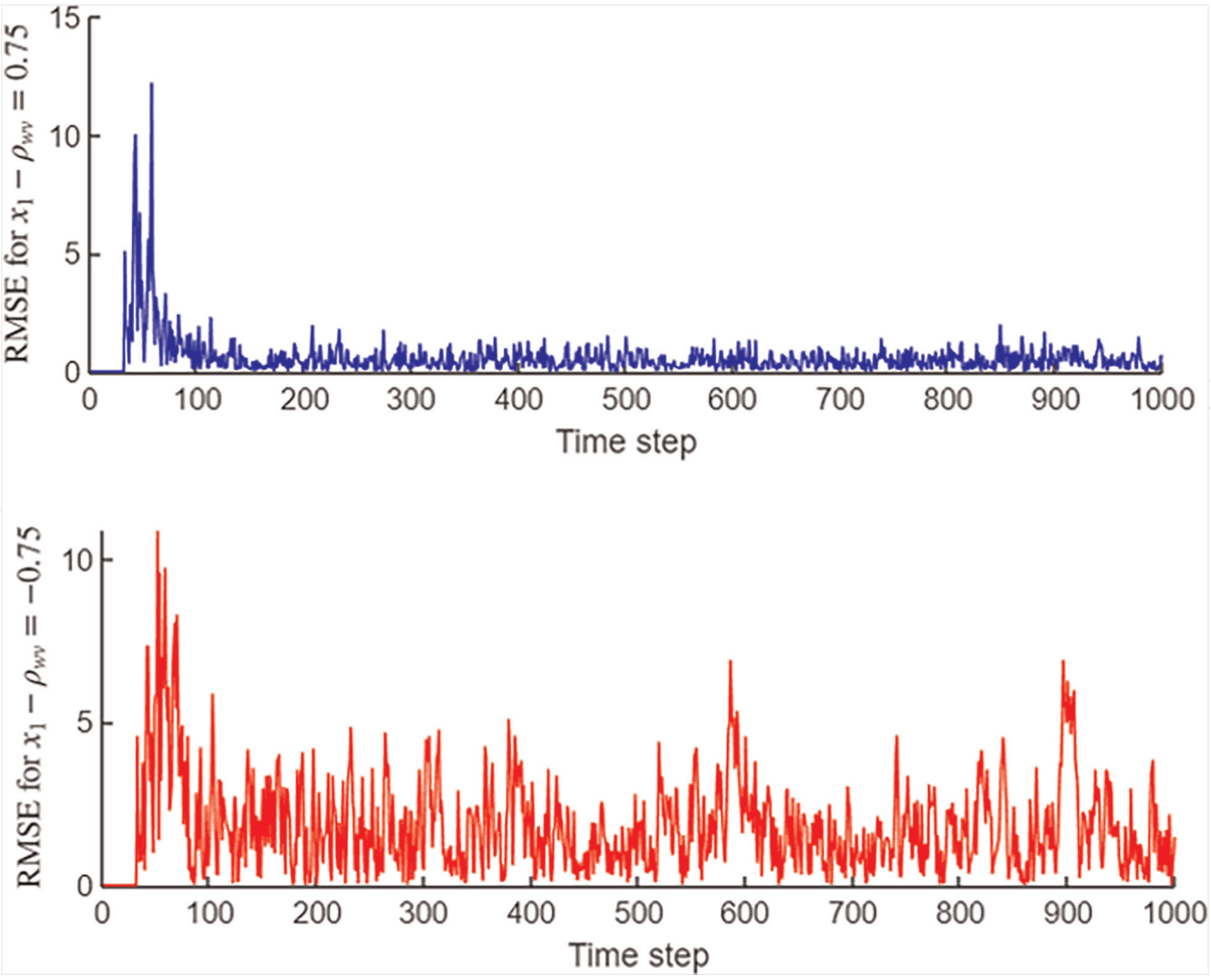

**Figure 13.**
*The root mean square error of the state x1 versus time.*

Concentrating on **Tables 5–8** and **Figures 11–14**, we can draw some conclusions from the table and figures.

- The parameter estimation errors produced by the KF-CN-RGELS algorithm decrease as the correlation coefficient increases in the positive direction. See **Figure 11** and **Table 5**.
- Under the same data length and the same correlation coefficient $\rho_{wv}$ the KF-CN-RGELS algorithm has a faster convergence rate than the SKF and AUG-KF algorithms – see **Figure 12** and **Tables 6** and 7.
- When the correlation coefficient is negative, the model and observations are less accurate. This affects the accuracy of the state estimation, which in turn affects the accuracy of the parameter estimation – see **Table 8** and **Figure 11** refer to Section 7 for details).
- The type of correlation between the process and measurement noises (increasing positively or negatively) affects the root mean-quare error of the system state. A highly positive correlation reduces the root mean square error, whereas a highly negative correlation increases the root mean square error of system states; see **Figure 13**.

Against $t\left(\rho_{w,v} = 0, 0.3, 0.5, 0.9, \text{ and } -0.7, \text{ for } R_v = 1.6, Q_w = [0.6{,}0.4]I_2\right)$ and Augmented KF algorithms $\left(\rho_{w,v} = 0.65 \text{ for } R_v = 1.6, Q_w = [0.6{,}0.4]I_2\right)$

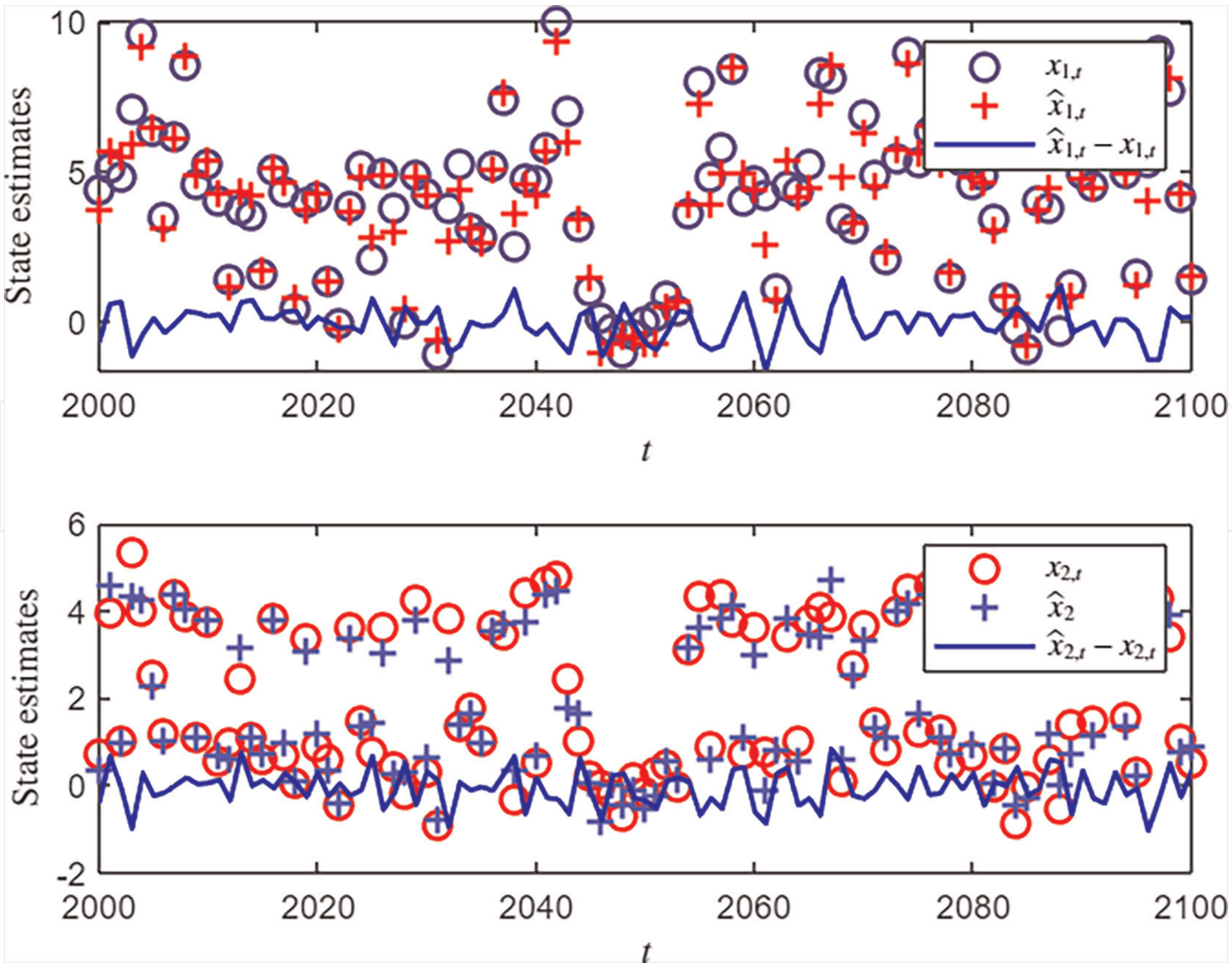

**Figure 14.**
*The true state, estimated state and state estimation error used in example 1, for* KF − CN − RGELS $(\rho_{w,v} = (0.75, -0.75))$ *for* $R_v = 1.6, Q_w = [0.6{,}0.4]I_2)$ $\rho_{w,v} = 0.75$.

## 9. Conclusion

The papers in this issue propose the KF-CN-RGELS algorithm to solve the combined state and parameter estimation problem in stochastic linear state-space systems perturbed by correlated processes and measurement noise. In this study, a known degree of correlation between measurement noise and process noise was theoretically predefined. To estimate the state of the system, the values of the correlation coefficients were chosen based on the considered real-time application situation. We developed a variant of the Kalman filter that considers noise correlations by adjusting the process noise matrix. Simulations show this new approach is successful, improving parameter estimation accuracy as noise correlation becomes more positive. The effectiveness of the proposed algorithm was tested theoretically by selecting several values for the positive and negative correlation coefficient. A negative correlation coefficient can have negative effects on the accuracy of both observations and models. It causes the measurements to be less reliable and introduces more uncertainty in determining the relationship between the observed data and the underlying process. This uncertainty, in turn, affects the filtering process that is used to estimate the true state and parameters from the observed data. As a result, the state and parameter estimates become less accurate. Many researchers in literature provide efficient ways to estimate the covariance matrices Q, R, and S that will be used in future work to guarantee best practice results. By estimating these matrices, the optimal correlation coefficient can be practically chosen, and according to Eq. (42), the algorithm gives satisfactory results from a practical point of view. Beyond its application to linear stochastic systems with KF-CN-RGELS, this approach opens doors to developing novel algorithms and tackling the identification of more complex systems like bilinear ones. Furthermore, the proposed KFCN-RGELS can be used to study real-time applications such as aircraft radar guidance systems, which is the case for cross-correlated noise systems.

## Conflict of interest

The authors declare no conflict of interest.

## Author details

Abd El Mageed Hag Elamin Khalid
College of Engineering, Al Neelain University, Khartoum, Sudan

*Address all correspondence to: elaminkhalid32@gmail.com